\documentclass[9pt,twoside]{rmaa-rho}
\RMxAAtemplatetype{\RMxAA} % Select your article type
\usepackage{aas_macros}

\vol{100}
\pages{1000-1006}
\thisyear{2026}
\doi{\href{https://doi.org/10.22201/ia.01851101p.20XX.XX.XX.XX}{https://doi.org/10.22201/ia.01851101p.20XX.XX.XX.XX}}

\title{A Homogenized Catalogue of Variable Stars in the Globular Cluster M22: Membership, Physical Parameters, and Distance}

\author[1,5,$\dagger$]{M. A. Yepez\orcidlink{0000-0001-8982-736X}}
\author[2]{B. Moreno-Amaro \orcidlink{0009-0003-1263-7420}}
\author[3]{M. Ornelas-Castañeda \orcidlink{0009-0009-6859-7716}}
\author[4]{A. Arellano Ferro \orcidlink{0000-0002-8514-2395}}
\author[1,5]{D. Deras\orcidlink{0000-0002-3730-9664}}

\affil[1]{Instituto Nacional de Astrof\'isica, \'Optica y Electr\'onica , Luis Enrique Erro No.1, 72840, Tonantzintla, Puebla, M\'exico}
\affil[2]{Benem\'erita Universidad Aut\'onoma de Puebla, Facultad de Ciencias F\'isico Matem\'aticas, Ciudad Universitaria, C.P. 72570, Puebla, M\'exico}
\affil[3]{Universidad de Guadalajara, Departamento de F\'isica, Revolución 1500, 44430 Guadalajara, Jalisco, M\'exico}
\affil[4]{Universidad Nacional Autónoma de México, Instituto de Astronomía, AP 70-264, CDMX 04510, México}
\affil[5]{Secretar\'ia de Ciencia, Humanidades, Tecnolog\'ia e Innovaci\'0n, Av. Insurgentes Sur 1582, 03940, Ciudad de M\'exico, M\'exico}
\leadauthor{Yepez et al.}
\smalltitle{Variable stars in M22}

\corres{M. A. Yepez}
\email{myepez@inaoep.mx}

\received{February 30th, 2026}
\accepted{\today}
\license{Texto de la licencia aquí}

\setbool{rho-abstract}{true} % Set false to hide the abstract
\setbool{rho-resumen}{true} % Set false to hide the abstract

\begin{abstract}
We present a photometric analysis of the variable stars in the field of the globular cluster M22 (NGC 6656), using time-series photometry from the literature and from \textit{OGLE} and \textit{Gaia}-DR3 data. From a sample of 666 variable stars in the field, we identify 94 as cluster members, including 10 RRab, 17 RRc, 2 Type II Cepheids, 15 SX Phe, 15 SR stars and 22 eclipsing binaries. Fourteen of these members are not included in the Catalogue of Variable Stars in Globular Clusters. Through the Fourier decomposition of the RR Lyrae light curves, we derived $[\mathrm{Fe/H}] = -1.73 \pm 0.17$ dex and $d = 3.27 \pm 0.14$ kpc. Among the RRc stars, V15 shows an additional frequency with a frequency ratio $f_{0.61}/f_\mathrm{1O} = 0.6152$, consistent with the RR$_{0.61}$ phenomenon. Using P-L relations for member SX Phe stars, we obtained $d = 3.31 \pm 0.17$ kpc, consistent with the RR Lyrae distance. We constructed a CMD comprising only cluster members, which allows us to confirm the evolutionary status of the cluster's RR Lyrae stars. None of the fitted isochrones satisfactorily reproduces the CMD morphology, which is consistent with the presence of multiple stellar populations in M22.

\end{abstract}


\begin{resumen}
Presentamos un análisis fotométrico de las estrellas variables en el campo del cúmulo globular M22 (NGC 6656), utilizando fotometría de series temporales proveniente de la literatura y de datos de \textit{OGLE} y \textit{Gaia}-DR3. A partir de una muestra de 666 estrellas variables en el campo, identificamos 94 como miembros del cúmulo, entre las que se encuentran 10 RRab, 17 RRc, 2 Cefeidas de tipo II, 15 SX Phe, 15 SR y 22 binarias eclipsantes. Catorce de estos miembros no están incluidos en el Catálogo de Estrellas Variables en Cúmulos Globulares. Por medio de la descomposición de Fourier de las curvas de luz de las estrellas RR Lyrae derivamos $[{\rm Fe/H}]=-1.73 \pm 0.17$ dex y $d=3.27 \pm 0.14$ kpc. Entre las estrellas RRc, V15 muestra una frecuencia adicional con razón de frecuencias $f_{0.61}/f_\mathrm{1O} = 0.6152$, consistente con el fenómeno RR$_{0.61}$. Mediante relaciones P-L para estrellas SX Phe miembros obtuvimos $d = 3.31 \pm 0.17$ kpc, consistente con los resultados de las estrellas RR Lyrae. Construimos un DCM sólo con estrellas miembros del cúmulo, lo que nos permite corroborar el estado evolutivo de sus estrellas variables. Ninguna de las isócronas ajustadas reproduce satisfactoriamente la morfología del CMD, lo cual es consistente con la presencia de múltiples poblaciones estelares en M22.

\end{resumen}

\begin{document}
\nolinenumbers
\maketitle
\pagestyle{fancy}\thispagestyle{firststyle}

%----------------------------------------------------------

\section{INTRODUCTION}

\label{sec:Intro}
\RMxAAstart{M}essier 22 (M22, NGC 6656) is a globular cluster located in the Sagittarius constellation ($\alpha = 18^\text{h}36^\text{m}23.94^\text{s}$, $\delta = -23^\circ54\text{'}17.1\text{''}$, J2000) at a distance of about 3.3 kpc \citep{Baumgardt2021}, with a [Fe/H]=-1.70 \citep{Harris1996}. Initially discovered by Abraham Ihle in 1665 and later cataloged by \cite{Messier1781}, M22 was originally classified as a nebula due to the limited resolving power of telescopes at the time. In 1783, William Herschel successfully resolved the cluster into its individual stellar components, establishing M22 as one of the first recorded globular clusters. Furthermore, it is the third closest globular cluster to the Sun, after M4 and NGC 6397. Its age is estimated at 12-14 Gyr \citep{VandenBerg2013}, making M22 an excellent object of study for ancient stellar populations and the early stages of the Milky Way.

Recent studies suggest that M22 is the result of the merger of two globular clusters  \citep{Lee2015}. This event likely occurred in a dwarf galaxy and was later accreted by our Galaxy. This could explain its considerable old age. Even more recently, evidence of as many as five distinct stellar subpopulations within the cluster has been reported \citep{Lee2020}. \citet{McKenzie2022} provide a comprehensive compilation of metallicity determinations spanning four decades in their Table 1, which consistently reports metallicity dispersions, and confirm with high-precision spectroscopy a $\Delta[\rm{Fe/H}] = 0.238$ dex between two populations. \citet{Lee2023} demonstrated this bimodality clearly through photometric CMDs, which distinctly separate the two metallicity populations with the same dispersion. This complexity further supports the idea of a complex evolutionary history and distinguishes M22 from the archetypal globular cluster. However, despite being one of the clusters closest to the Sun, factors such as high-field star contamination, complicate the study of M22, making of it a challenging object for investigating its variable stars. In addition, the cluster is affected by significant differential reddening \citep{Marino2009}. Several studies have adopted different values for the reddening, ranging from average estimates, such as $E(B-V) = 0.34$ \citep{Harris1996}, to those with reddening ranges as large as $0.3 < E(B-V) < 0.5$ \citep{Marino2009}. This peculiarity has probably contributed to exclude this cluster from certain studies related to the chemical composition of stellar populations \citep[e.g.,][]{Carreta2010}.

Despite these peculiarities, photometric studies have revealed that the cluster exhibits an extended horizontal branch morphology \citep{Monaco2004}, with a well-defined instability strip populated by numerous RR Lyrae stars \citep{Rozyczka2017} (hereafter R17). The mean pulsation periods of these variables place M22 within the Oosterhoff type II class. Their pulsational properties, therefore, provide valuable insights about the evolutionary effects along its HB.

While numerous studies have focused on the global properties of M22, most investigations have primarily addressed population analyses, average chemical distributions, and collective dynamical properties. In contrast, no study has yet carried out a detailed characterization of the physical parameters of its individual stars. The work by \cite{Kunder2013a} represents the closest effort in this direction, providing average metallicities, distances, and magnitudes for the RR Lyrae population in M22. Nevertheless, the systematic determination of the physical parameters of the RR Lyrae member stars presented in this work is essential for a deeper understanding of the system's evolutionary state.

M22 has a large variable star population; the first variables, V1-V16, were discovered by \cite{Bailey1902}, who also carried out one of the earliest classifications of RR Lyrae subtypes, while V17 was discovered by \cite{Shapley1927}. \cite{Sawyer1944} later expanded the catalog with the variables V18–V25 and subsequently incorporated V26–V31, originally detected by \cite{Hoffleit1972}. Variables V32–V35 were identified between the 1970s and 1980s through the work of \cite{Wehlau1977,Wehlau1978}, and \cite{Lloyd1978}. In the following decades, photometric studies significantly increased the number of known variables: \cite{Kravtsov1994} discovered V36–V43, while \cite{Kaluzny2001} identified dozens of new variables as part of the Cluster AgeS Experiment (CASE) project. Subsequent studies by \cite{Pietrukowicz2005}, \cite{Kunder2013a} and \cite{Sahay2014} added RR Lyrae stars, long-period variables, and cataclysmic systems associated with the cluster. 

In the study of R17, they report light curves for 359 variable stars, 238 of which are new detections. They classified 102 variables as members or probable members. A year later, \citet{Rozyczka2018} (hereafter R18) used the friends-of-friends method (FoF) to detect transient or aperiodic phenomena in large photometric databases. They reported 31 variables or suspected variables, but only 2 are cataloged as cluster members. These authors labeled the variables by their membership: $N$ for field stars and $U$ for unclassified variables.

According to R17 and R18, there are 390 variable stars in the field of view of M22, 105 of which are cataloged as members. However, \citet{Prudil2024} report that only 86 of these stars are actual members.

More recently, \citet{Alonso-Garcia2021} conducted a search for variable stars in M22 using near-infrared multi-epoch photometry from the VVV survey, with membership assignment based on proper motions from the VIRAC2 catalog. They reported 38 variable stars as cluster members, of which ten are new discoveries: C5, C19, C20, C29, C200, C205, C216, C263, C295, and C359. Among the previously known members, no new RR Lyrae stars were identified.

This suggests an uncertainty in the membership assignment and underscores the need for a more comprehensive re-evaluation, integrating multiple photometric catalogs to construct a homogeneous database of variable stars in M22, with the aim of refining period determinations, improving classifications and reassessing membership. 

In this paper, we present a study of the variable stars in the field of M22, using the Optical Gravitational Lensing Experiment (OGLE) database \citep{Udalski1992,Udalski2015} and $Gaia$-DR3 \citep{Gaia2016b,Gaia2023j} surveys and the CASE database. A total of 666 variable stars were analyzed, including both members and non-members. We estimated the physical parameters of each RR Lyrae star using their light curve Fourier decomposition. From the $Gaia$-DR3 data, transformed to $V$ and $I$ of the Johnson-Cousins photometric system \citep{Gaia_Documentation2022}, we constructed a color-magnitude diagram (CMD) and overplotted isochrones and Zero Age Horizontal Branch (ZAHB) according to the age reported in the literature. We also analyzed in detail some peculiar stars.

The paper is structured as follows: Section~\ref{sec:Data} details the data acquisition. In Section~\ref{sec:Vars}, we discuss the variable stars and their membership. In Section~\ref{sec:reddening}, we detail the technique used to correct for differential reddening in the cluster. Section~\ref{sec:Four} shows the physical parameters derived for the RR Lyrae stars. In Section~\ref{sec:cmd}, we present the final CMD. In Section~\ref{sec:BD}, we provide a distance estimate based on SX Phe stars that are members of M22. In Section~\ref{sec:SXPL}, we present the Bailey diagram for all the RR Lyrae stars in the field. Finally, in Section \ref{sec:conclusions}, we give our conclusions.

\section{Data Compilation}
\label{sec:Data}

We used four datasets to identify light curves of variable stars in the field of M22. First, we used data from R17 and R18. These data are of very high quality and are available through the CASE database; however, not all detected light curves are publicly accessible. Through private communication, Dr. Rozyczka kindly provided the missing light curves. A total of 391 light curves of variable stars were obtained from this source within a radius of 12.4$\arcsec$. All curves are in the Johnson $V$ passband.

The second set of light curves is from the OGLE database. We searched for variable stars within a radius of 22 arcmin, which is the radius at which stars cease to be considered members of M22 according to the membership analysis of \citet{Vasiliev2021}. From the OGLE database, 357 light curves were recovered.

The third source of light curves was the $Gaia$-DR3 database. Using the same central coordinates and tidal radius as for OGLE, we selected sources flagged as variable. To complement the OGLE and CASE data sets, we cross-matched their coordinates with $Gaia$-DR3 variable sources. Once downloaded, the $Gaia$-DR3 photometry was transformed into the Johnson system using the transformation equations provided by $Gaia$-DR3 documentation \citep{Gaia_Documentation2022}. This resulted in 218 light curves.

Finally, the fourth set includes the ten new cluster members reported by \citet{Alonso-Garcia2025}. We searched for these stars in the OGLE and Gaia-DR3 databases and found light curves for only four of them: C29, C200, C205, and C295. Among these, C29 was previously identified as U39 by R17, and C295 lies outside the 22 arcmin region considered in this work.

Once all data were collected, we cross-matched the above data sets and, when applicable, combined the light curves for each variable star. A total of 666 light curves of variable stars were recovered in the field of view of M22. For each light curve, we refined the periods using both the string-length method \citep{Burke1970, Dworetsky1983} and the software \textsc{Period04} \citep{Lenz2005}, adopting the best fit result from both techniques. The epochs of maximum light were identified directly from each light curve. It is worth noting that of the 140 variables listed in the May 2018 edition of the Catalogue of Variable Stars in Globular Clusters (CVSGC) \citep{Clement2001}, the following stars were not found in any the above soucers, hence no data is available to us: V28, PK-04, PK-06, PK-07, PK-08, PK-09, PK-10, PK-11, SLW-6, SLW-7 and SLW-9.

\begin{table*}[htbp]
	\Centering
	\caption{Cross-identification of variable stars in the field of M22.}
	\label{tab:crossid}
	\footnotesize
	\begin{tabular}{lcc|lcc|lcc}
		\toprule
		\textbf{ID} & \textbf{OGLE} & \textbf{$Gaia$-DR3} & \textbf{ID} & \textbf{OGLE} & \textbf{$Gaia$-DR3} & \textbf{ID} & \textbf{OGLE} & \textbf{$Gaia$-DR3} \\
		\midrule
V1         & RRLYR-36670 & 4077588422733581056  & KT-08      & ECL-423208  & --                   & V108       & --          & 4077492863921299840  \\
V2         & RRLYR-36681 & 4077588147768728192  & KT-10      & --          & 4077587735451582848  & V109       & --          & 4077497747325912064  \\
V3         & --          & 4077593881548557312* & KT-12      & RRLYR-36679 & 4077588525812691328  & V110       & --          & 4077593714050152192  \\
V4         & RRLYR-36673 & 4077588349639105408* & KT-13      & ECL-423192  & --                   & V111       & --          & --                   \\
V5         & --          & 4077494758098041088* & KT-14      & --          & 4077588491366194560* & V112       & DSCT-09968  & --                   \\
V6         & RRLYR-36669 & 4077494478834899840* & KT-15      & ECL-423188  & 4077589281649574784* & V113       & --          & 4077588349639106944  \\
V7         & --          & 4077593125634310144  & KT-16      & ELL-025311  & 4077587318838435968  & V114       & --          & 4076742314152851456  \\
V8         & --          & 4077494517579748480  & KT-18      & ECL-423184  & 4077589281653651840  & V115       & --          & 4076742692133304192  \\
V9         & --          & 4077494719377498240* & KT-20      & ECL-423175  & 4077589178562141568* & V116       & --          & --                   \\
V10        & --          & 4077493727234716928* & KT-23      & ECL-423159  & 4077588869324708864* & V117       & --          & 4077494483220024704  \\
V11        & --          & 4077588452718318080* & KT-26      & --          & 4077588452718411904  & V118       & ECL-423093  & 4077592958236095104  \\
V12        & RRLYR-36674 & 4077587598099748224* & KT-27      & DSCT-09966  & 4077588353927463808  & V119       & --          & --                   \\
V13        & RRLYR-36677 & 4077589174272939392* & KT-28      & DSCT-09964  & 4077588830675528960  & V120       & --          & --                   \\
V14        & --          & 4077594023371124224* & KT-29      & DSCT-09963  & --                   & V121       & --          & --                   \\
V15        & RRLYR-36680 & 4077587559356597888* & KT-33      & --          & --                   & V122       & ECL-423187  & --                   \\
V16        & RRLYR-36682 & 4077587696804128768* & KT-34      & DSCT-09961  & 4077494586231991680  & V123       & --          & 4077588903683057664  \\
V17        & --          & 4077498739434847744* & KT-36      & RRLYR-36667 & 4077494547554369152* & V124       & --          & --                   \\
V18        & RRLYR-36668 & 4077593331792623616* & KT-37      & RRLYR-36665 & 4077494856782483072  & V125       & --          & 4077593130018233984  \\
V19        & RRLYR-36671 & 4077588830675524352* & KT-38      & --          & 4077494410107507072  & V126       & --          & --                   \\
V20        & RRLYR-36666 & 4077494375746153600* & KT-39      & ECL-423107  & 4077591716884964992  & V127       & --          & 4077590346878801024  \\
V21        & RRLYR-36675 & 4077588560172542848* & KT-40      & ECL-423102  & 4077591927427161472* & V128       & --          & 4077587456285957504  \\
V22        & RRLYR-36640 & 4077510009429113088* & KT-41      & ECL-423099  & --                   & V129       & --          & 4077588834965053824  \\
V23        & RRLYR-36672 & 4077588349639104256  & KT-42      & ECL-423203  & 4077588903770488448  & V130       & --          & 4077493624155503104  \\
V24        & T2CEP-0927  & 4077588418358577024* & KT-43      & ECL-423166  & 4077493761594261248  & V131       & --          & --                   \\
V26        & --          & 4077511628731690496* & KT-45      & DSCT-09965  & 4077588422647166336  & V132       & --          & 4077493692945912320  \\
V27        & RRLYR-36648 & 4077512242812406528* & KT-46      & ECL-423150  & 4077493555435414784  & V133       & --          & 4077588147768329856  \\
V29        & RRLYR-36678 & 4077595810077798528* & KT-48      & --          & --                   & V134       & ECL-423136  & 4077494551868848128  \\
V36        & RRLYR-36651 & 4077498262791669248  & KT-51      & --          & 4077588143480811392  & V135       & --          & 4076837245778120576* \\
V37        & ECL-423040  & 4077606805196159232* & KT-54      & DSCT-09967  & 4077589242983815936  & V136\_R    & --          & 4077493619841438464  \\
V38        & RRLYR-36657 & 4077606564677584384  & KT-55      & --          & 4077588452718412288* & V136\_f    & --          & 4077588177831883520  \\
V25        & RRLYR-36683 & 4077590754812727296* & PK-05      & --          & --                   & V137       & --          & --                   \\
V30        & --          & 4076836592917559040* & Ku-1       & --          & 4077494277061730048* & V138       & --          & 4077588766330992384  \\
V31        & --          & 4076740390030815488* & Ku-2       & RRLYR-36662 & 4077592747677139328* & V139       & --          & 4077494792458231680  \\
V32        & --          & 4077495136055282176* & Ku-3       & --          & 4076742554671731712* & V140       & --          & 4077588697611640192  \\
V33        & --          & 4076739324878814464* & Ku-4       & ELL-025312  & 4077589384806381952  & V141(U01)  & --          & 4077592232281074688  \\
V34        & --          & 4077587598099722624  & CV1        & --          & --                   & V142(U08)  & --          & 4077592374018441216  \\
V35        & --          & 4077588388373791616* & CV2        & --          & --                   & V143(U33)  & --          & 4077493349276734080  \\
V39        & --          & 4077493177478762880* & P1         & --          & --                   & V144(U56)  & --          & 4077589075482041984  \\
V40        & --          & 4077593885932174720* & SLW-4      & --          & 4077494788072543104* & V145(U64)  & --          & 4077589178581183616  \\
V41        & --          & 4076838860660651776* & SLW-5      & --          & 4077588628892060160  & V146(U65)  & --          & 4077592923859669504* \\
V42        & --          & 4077591137152796032* & SLW-8      & --          & 4077494513194647168  & V147       & DSCT-09970  & 4077613707117780864  \\
V43        & --          & 4076838865031338368* & SLW-11     & --          & 4077588491453012992  & V148       & DSCT-15910  & 4076838620142445440  \\
KT-01      & --          & --                   & V102       & --          & 4077588079048829568  & V149       & ELL-025305  & 4077499053065707392  \\
KT-02      & ECL-423217  & 4076836803447045376  & V103       & --          & 4077588525734757248  & V150?(U41) & --          & 4077498022273396480  \\
KT-03      & ECL-423216  & 4077588285225930496  & V104       & --          & --                   & V151(C200) & ECL-422846  & --                   \\
KT-04      & --          & 4077588182128332928  & V105       & --          & 4077494517509048448  & V152(C205) & ECL-423378  & --                   \\
KT-05      & --          & 4077589144219360512  & V106       & --          & 4077591751244710656  & V153(C295) & ECL-423176  & 4077618792366741504* \\
KT-07      & ECL-423211  & 4077587288773795200* & V107       & --          & 4077587598012702208  & C29(U39)   & ECL-423130  & 4077493589795539072* \\
		\bottomrule
	\end{tabular}
	\tabletext{Note: ID in parentheses corresponds to an alternative designation found in the literature: prefix U refers to the catalogue of R17, and prefix C to that of \citet{Alonso-Garcia2021}. OGLE ID shown without the ``OGLE-BLG-'' prefix. An asterisk (*) in the $Gaia$-DR3 column indicates that the star has a variability flag in the $Gaia$-DR3 database. The whole table is in the CDS database.}
\end{table*}

\section{Variable stars in M22}
\label{sec:Vars}

\subsection{Overview}
M22 has a large population of variable stars. The CVSGC, lists 140 variables of a variety of types: RRab and RRc, SX Phe, SR, type II Cepheids, eclipsing binaries, Mira (M) and long period irregulars (L), dwarf novae (UG) and extrinsic or rotating BY Dra stars. 

Only 86 of the 140 variables in the CVSGC have been labeled as likely cluster members \citep{Prudil2024}. We identified 13 variables in our general data that are cluster members \citep{Vasiliev2021,Alonso-Garcia2021}, and as such we suggest to label them with the prefix `V' employed in the CVSGC from V141 to V153 (see Table \ref{tab:crossid}). In the present paper, for the light curve morphology analysis and the employment of RR Lyrae stars as physical parameter indicators, we shall concentrate on these member stars, see  Table \ref{tab:datosgenerales1}. The light curves data, periods, amplitudes and coordinates shall be made available in the Centre de Données astronomiques de Strasbourg (CDS).

Along our detaled analysis of individual stars, a number of them showed some peculiarities. Those cases are discussed in detail in Appendix \ref{sec: ap-A}.

\subsection{On the membership of variable stars}
\label{sec:membership}

One of the main challenges in the characterization of the CMD of a globular cluster is the stellar membership. If a proper analysis is performed, then we can have a cleaner CMD whose morphology is only sculpted by stellar evolution. The membership status is of particular relevance for variable stars if the aim is to use some variable types as indicators of physical parameters of the parental cluster, such as distance and metallicity. It is not possible to derive meaningful mean physical parameters for the cluster from its variable stars population if their membership status is not considered. For this purpose, we relied on previous membership analysis in the field of M22, specifically that of \citet{Vasiliev2021}, which had already been used for stars in the CVSGC by \citet{Prudil2024}.

By cross-matching our large catalogue of 666 variable stars with the members listed by R17, R18, \cite{Alonso-Garcia2021}, \citet{Prudil2024} and \citet{Vasiliev2021}, we identified 94 variable stars as members of M22. Among them there are 10 RRab, 17 RRc, 2 Type II Cepheids, 15 SX Phe, 15 SR, 19 Eclipsing binaries and 13 of other types showing periodic or sinusoidal variation probably of rotational origin.

We note that in the list of 70 stars labeled as ``U'' (unclassified) by R17, we found 7 that are cluster members according to the analysis of \citet{Vasiliev2021} (1 SX\,Phe, 3 EW, 1 SR, 1 Per and 1 Sin). Among the 4 new members found in \citet{Alonso-Garcia2021}, one corresponds to a star also listed as unclassified in R17 (U39,
classified as EA), bringing the total number of previously unclassified R17 stars confirmed as members to 8. We also found three new members from the OGLE database (2 SX Phe and 1 ECL). All these new member variables are already included in the grand total of 94 member variables in the field of M22. A summary of the general data for these variables is given in Table \ref{tab:datosgenerales1}. Comments on peculiar cases, such as V136\_R, V136\_f, N107, and others, will be discussed in the appendix \ref{sec: ap-A}.

\begin{figure*}[ht!]
	\begin{center}
		\includegraphics[width=17.5cm]{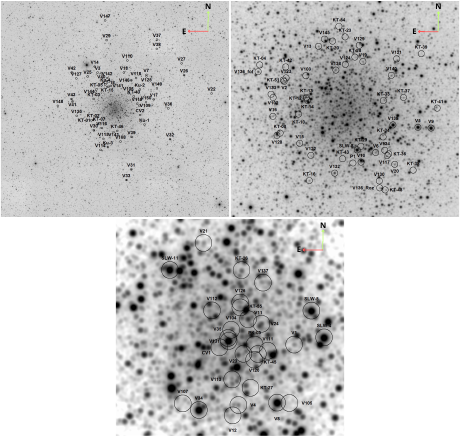}
		\caption{Maps of variable stars in the field of view of M22. Top: The left panel shows the full field with a radius of 22 arcmin, while the right panel corresponds to a zoom of the central region within a radius of 7 arcmin. Bottom: The panel shows the innermost zone with a radius of 3.5 arcmin. The background image was retrieved from the 2MASS $H$-band survey (1.66~$\mu$m) using the CDS HiPS2FITS service. Only the variable stars listed in Table \ref{tab:datosgenerales1} are shown.}
		\label{Fig:Chart_id}
	\end{center}
\end{figure*}

In this paper, we present a total of 144 variable stars: 94 members, 36 field and 14 unclassified, all of which are shown in the ID chart (Fig. \ref{Fig:Chart_id}). The whole collection has another 420 fields and 102 unclassified variable stars. Those variables that had not been previously identified and were directly detected by OGLE have been relabeled with the letter $O$ followed by a number.

\begin{figure*}[htbp]
	\begin{center}
		\includegraphics[width=16cm]{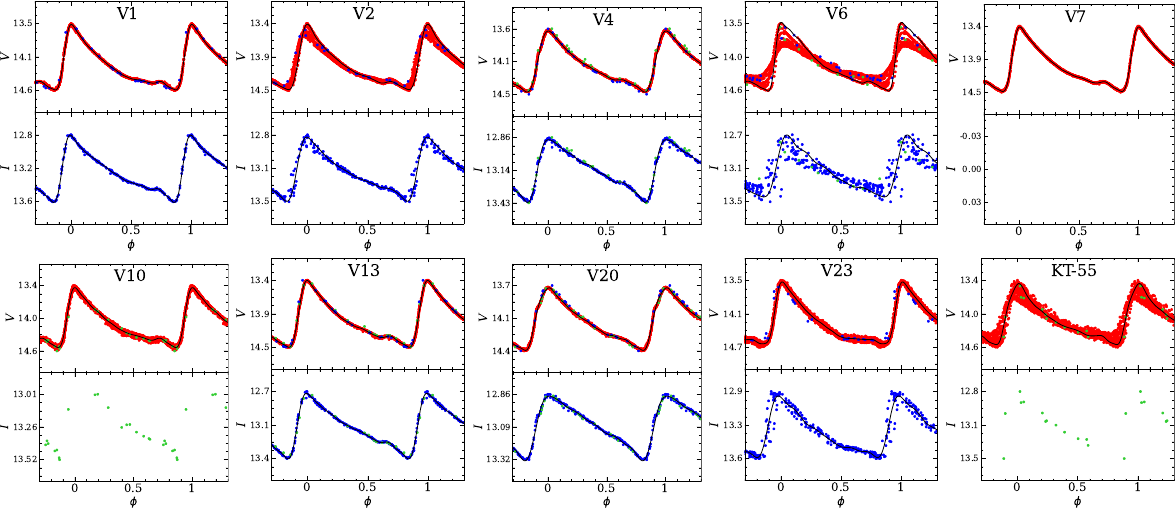}
		\caption{Light curves of RRab Stars. Red, blue and green points are data from R17 or R18, OGLE and $Gaia$, respectively. \\ The black solid line is the Fourier fit.}
		\label{Fig:Vars_RRab}
	\end{center}
\end{figure*}

\begin{figure*}[htbp]
	\begin{center}
		\includegraphics[width=16cm]{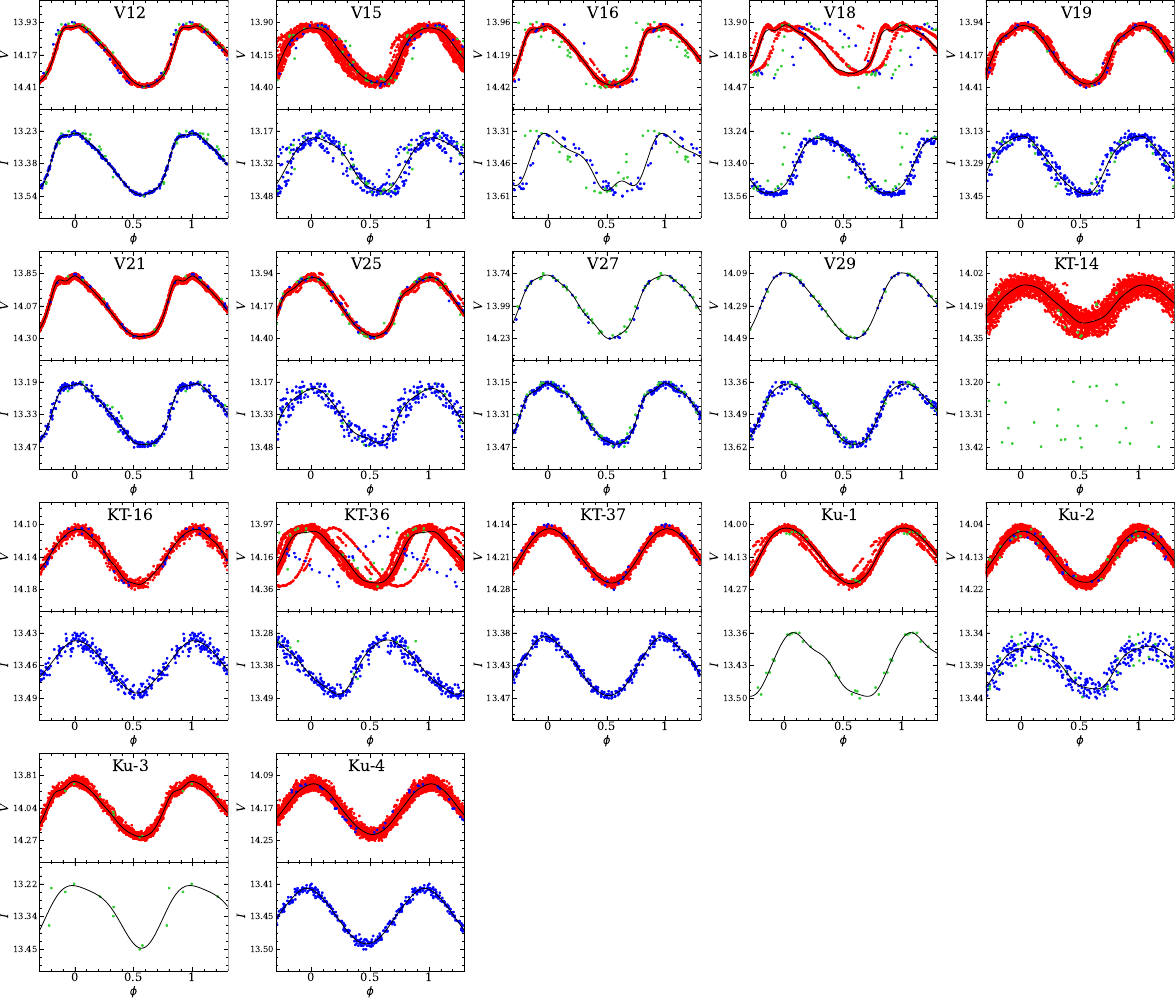}
		\caption{Light curves of RRc Stars. The color code is as in Fig. \ref{Fig:Vars_RRab}.}
		\label{Fig:Vars_RRc}
	\end{center}
\end{figure*}

\begin{figure*}[htbp]
	\begin{center}
		\includegraphics[width=16cm]{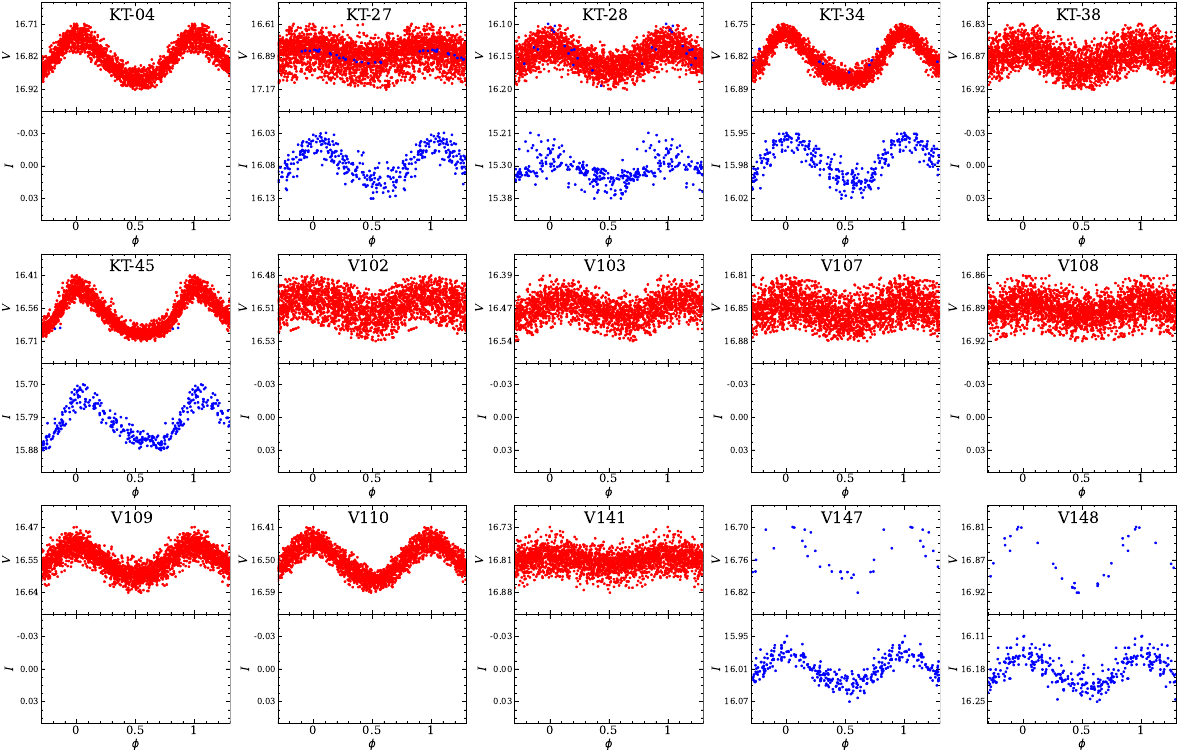}
		\caption{Light curves of SX Phe Stars. The color code is as in Fig. \ref{Fig:Vars_RRab}.}
		\label{Fig:Vars_SX_Phe}
	\end{center}
\end{figure*}

\begin{figure*}[htbp]
	\begin{center}
		\includegraphics[width=16cm]{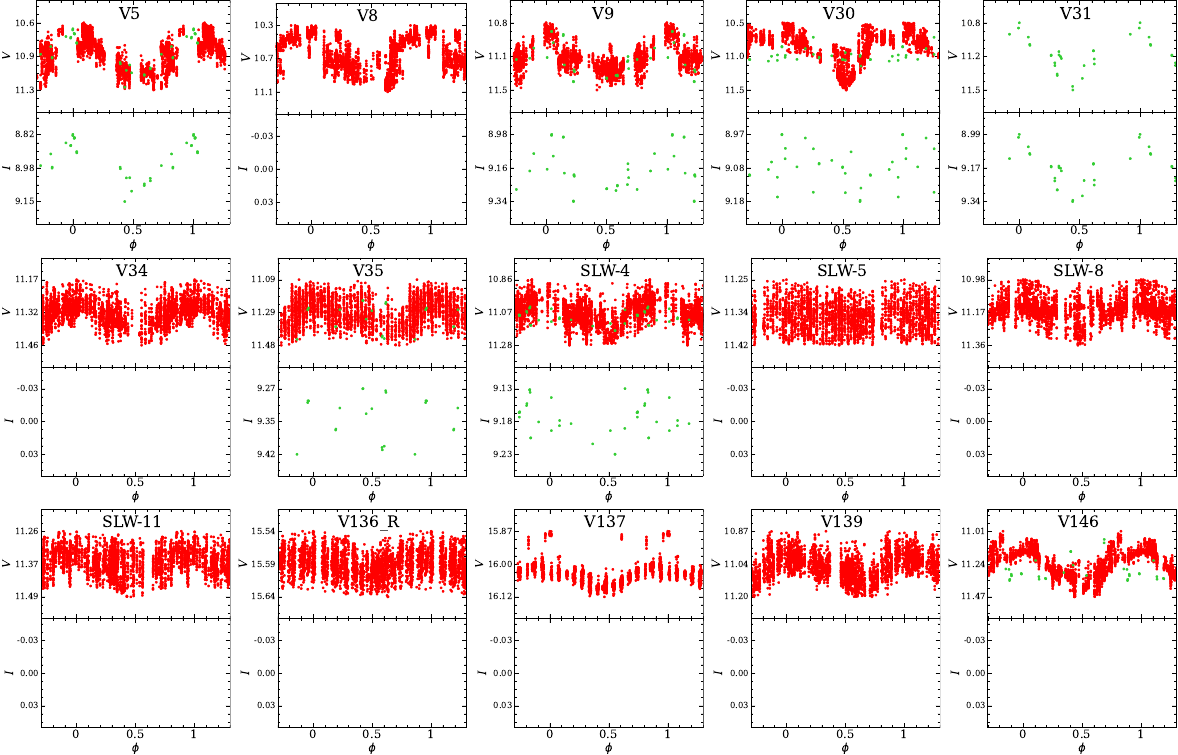}
		\caption{Light curves of SR Stars. The color code is as in Fig. \ref{Fig:Vars_RRab}.}
		\label{Fig:Vars_SR}
	\end{center}
\end{figure*}

\begin{figure}[htbp]
	\begin{center}
		\includegraphics[width=8.5cm]{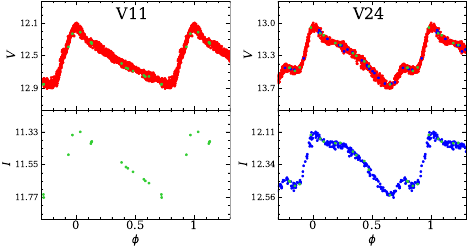}
		\caption{Light curves of V11 and V24, both BL Her Stars. The color code is like Fig \ref{Fig:Vars_RRab}.}
		\label{Fig:Vars_CW}
	\end{center}
\end{figure}

\begin{figure*}[htbp]
	\begin{center}
		\includegraphics[width=16cm]{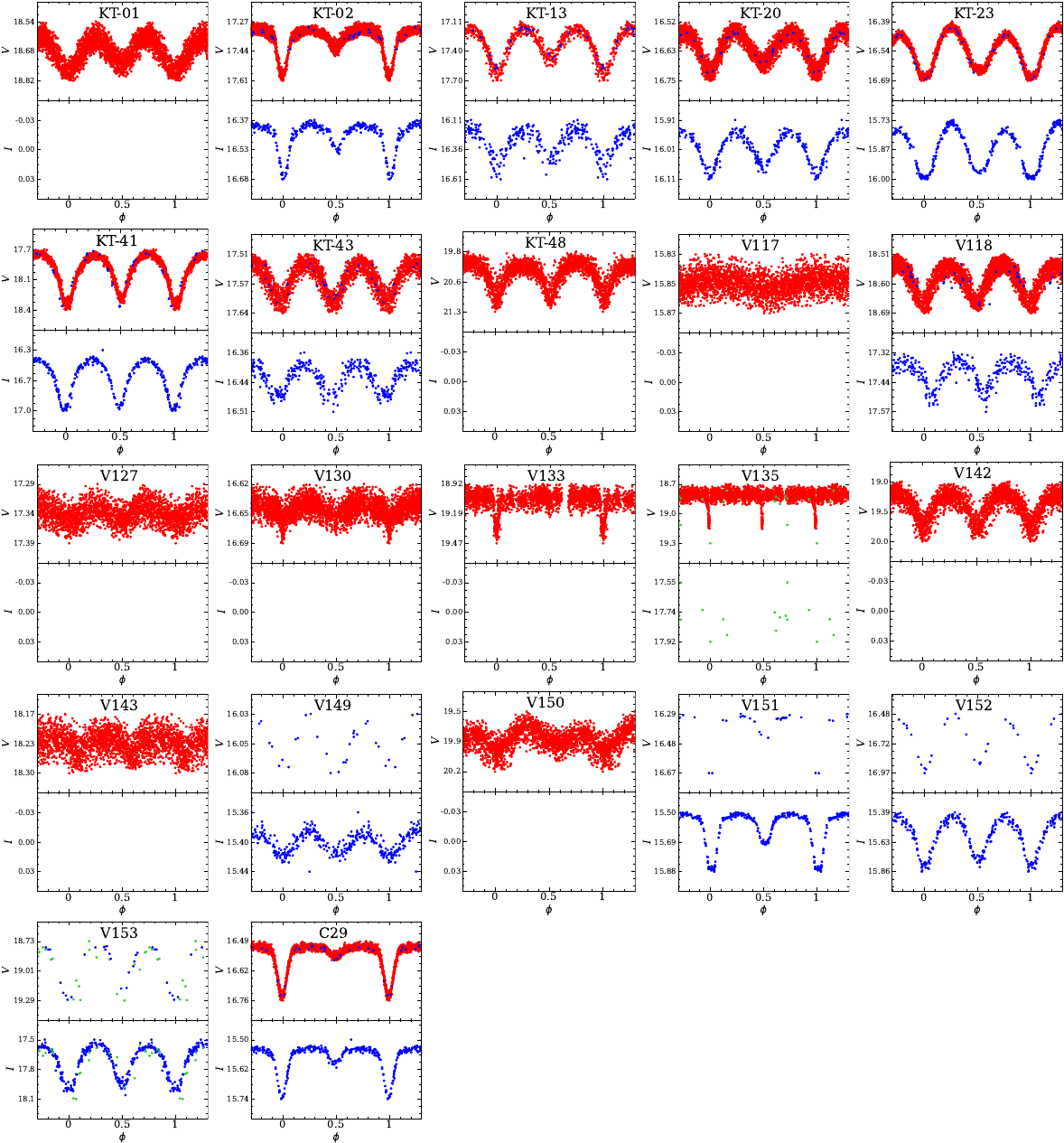}
		\caption{Light curves of Eclipsing binary stars. The color code is like Fig. \ref{Fig:Vars_RRab}.}
		\label{Fig:Vars_ECL}
	\end{center}
\end{figure*}

\begin{figure*}[htbp]
	\begin{center}
		\includegraphics[width=16cm]{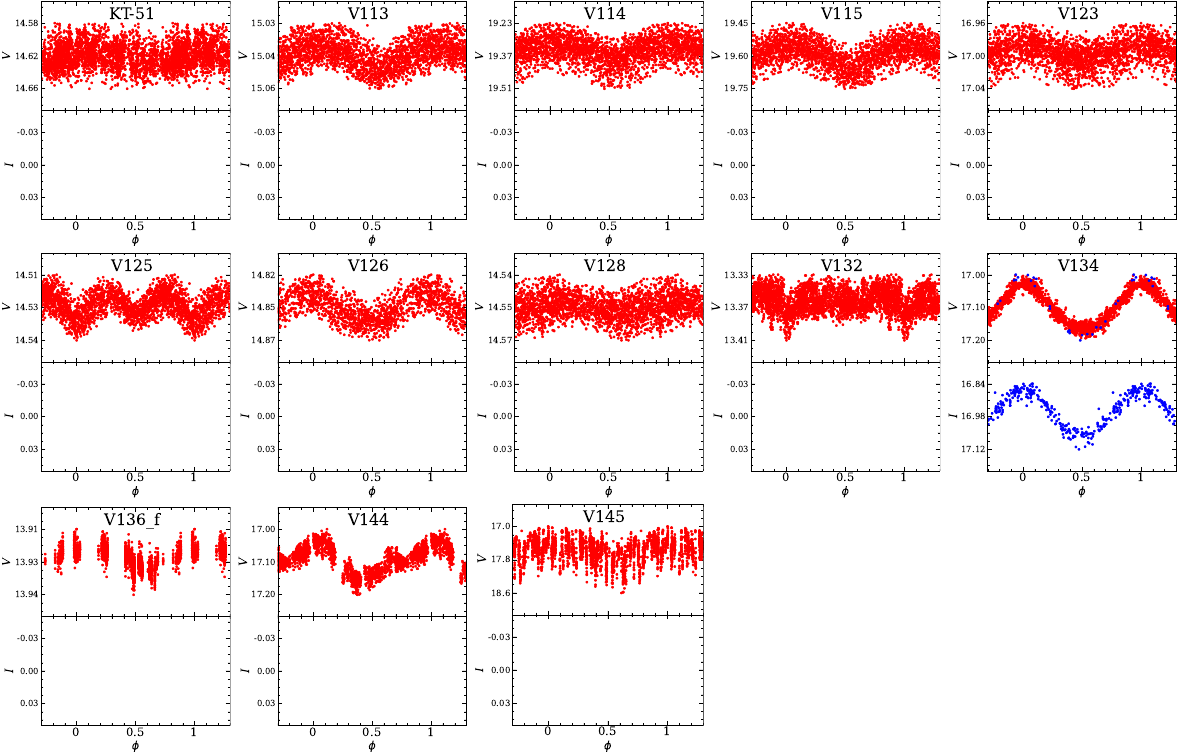}
		\caption{Light curves of stars classified as sin (sinusoidal), per (periodic), susp (suspected), or v (variable). The color code is like Fig. \ref{Fig:Vars_RRab}.}
		\label{Fig:Vars_Ext}
	\end{center}
\end{figure*}

In the Figs \ref{Fig:Vars_RRab}, \ref{Fig:Vars_RRc}, \ref{Fig:Vars_SX_Phe}, \ref{Fig:Vars_SR}, \ref{Fig:Vars_CW}, \ref{Fig:Vars_ECL} and \ref{Fig:Vars_Ext} we show the light curves of all member stars for which we have light curves available. Figures are organized by variable types.

\begin{table*}[htbp]
	\Centering
	\caption{General Data of Variables in the Field of M22.}
	\label{tab:datosgenerales1}
	\footnotesize
	\begin{tabular}{llrrrrrrlll}
		\toprule
		\textbf{Variable} & \textbf{Type} & \textbf{$<V>$} & \textbf{$<I>$} & \textbf{$A_{V}$} & \textbf{$A_{I}$} & \textbf{Period} & \textbf{$HJD_{max}$} & \textbf{RA} & \textbf{Dec.} & \textbf{Memb.} \\
		&               & (mag)          & (mag)          & (mag)            & (mag)            & (days)          & (d+2450000)          & (J2000.0)   & (J2000.0)     & (M/F/U) \\
		\midrule
		V1        & RRab   & 14.247 & 13.247 & 1.087 & 0.740 & 0.615543    & 4293.7416 & 18:36:19.56 & -23:54:32.9 & M \\
		V2        & RRabBl & 14.071 & 13.190 & 1.112 & 0.739 & 0.641721    & 1784.5959 & 18:36:34.91 & -23:53:06.5 & M \\
		{V3}        & RRabBl & 15.781 & \textit{14.831}  & 1.366 & 0.613 & 0.515444    & 1772.4824 & 18:36:38.00 & -23:47:15.3 & F \\
		V4        & RRab   & 14.113 & 13.140 & 0.845 & 0.563 & 0.716396    & 6195.5132 & 18:36:23.36 & -23:55:29.3 & M \\
		V5        & SR     & 10.937 & 8.988  & 0.692 & 0.330 & 92.88   & 7323.4674 & 18:36:10.58 & -23:55:00.6 & M \\
		V6        & RRabBl & 14.136 & 13.126 & 1.103 & 0.824 & 0.638489    & 1675.9161 & 18:36:18.25 & -23:56:03.6 & M \\
		{V7}        & RRab   & 14.062 & \textit{13.206}  & 1.144 & -- & 0.649520    & 1725.6668 & 18:35:57.60 & -23:47:40.8 & M \\
		{V8}        & SR     & 10.673 & \textit{13.206}  & 0.778 & -- & 58.48   & 1772.5547 & 18:36:20.72 & -23:55:27.1 & M \\
		V9        & SR     & 11.134 & 9.066  & 0.688 & 0.355 & 91.71   & 2082.8905 & 18:36:08.20 & -23:55:02.9 & M \\
		{V10}      & RRab   & 14.115 & \textit{13.247} & 1.183 & 0.516 & 0.646027    & 2133.5249 & 18:36:20.93 & -23:56:27.3 & M \\
		V11       & BLHer   & 12.530 & 11.556 & 0.839 & 0.446 & 1.690486    & 2104.6499 & 18:36:22.71 & -23:54:09.1 & M \\
		V12       & RRc    & 14.179 & 13.386 & 0.477 & 0.307 & 0.322623    & 5299.9188 & 18:36:23.76 & -23:55:39.0 & M \\
		V13       & RRab   & 14.059 & 13.113 & 1.063 & 0.726 & 0.672531    & 1772.6040 & 18:36:28.65 & -23:51:39.7 & M \\
		{V14}       & M      & 14.823 & \textit{10.767} & 5.151 & 2.496 & 200.0  & 1695.8604 & 18:36:40.51 & -23:46:07.4 & F \\
		V15       & RRc$_{0.61}$ & 14.145 & 13.321 & 0.496 & 0.308 & 0.373201    & 4294.6941 & 18:36:32.08 & -23:55:40.9 & M \\
		V16       & RRcBl    & 14.193 & 13.463 & 0.462 & 0.302 & 0.325293    & 2072.7844 & 18:36:36.97 & -23:54:33.5 & M \\
		V17       & SR     & 14.047 & 10.886 & 1.355 & 0.649 & 137.28  & 7881.7751 & 18:35:51.37 & -23:52:24.2 & F \\
		V18       & RRcBl  & 14.131 & 13.402 & 0.573 & 0.321 & 0.324830    & 1692.7948 & 18:36:16.15 & -23:47:15.1 & M \\
		V19       & RRc    & 14.170 & 13.296 & 0.473 & 0.318 & 0.384027    & 6394.8510 & 18:36:20.59 & -23:52:17.4 & M \\
		V20       & RRab   & 14.105 & 13.078 & 0.683 & 0.458 & 0.756134    & 1786.5291 & 18:36:14.80 & -23:56:33.1 & M \\
		V21       & RRc    & 14.080 & 13.328 & 0.452 & 0.280 & 0.327135    & 4293.7416 & 18:36:25.75 & -23:52:58.1 & M \\
		V22       & RRab   & 14.434 & 13.559 & 1.074 & 0.643 & 0.624535    & 7000.1089 & 18:35:03.46 & -23:51:15.5 & F \\
		V23       & RRab   & 14.279 & 13.351 & 1.281 & 0.754 & 0.551621    & 1693.7650 & 18:36:23.02 & -23:54:41.6 & M \\
		V24       & BLHer  & 13.375 & 12.337 & 0.727 & 0.453 & 1.714854    & 1756.6564 & 18:36:21.79 & -23:54:13.4 & M \\
		V25       & RRc    & 14.172 & 13.330 & 0.467 & 0.308 & 0.402378    & 1762.7570 & 18:36:46.29 & -23:48:02.4 & M \\
		V26       & SR     & 15.715 & 12.122 & 3.675 & 2.087 & 309.00  & 7456.2530 & 18:35:26.30 & -23:47:45.8 & F \\
		V27       & RRc    & 13.985 & 13.305 & 0.494 & 0.315 & 0.342784    & 5490.4986 & 18:35:28.34 & -23:45:29.9 & M \\
		V29       & RRc    & 14.285 & 13.490 & 0.406 & 0.260 & 0.304265    & 6424.8809 & 18:36:30.60 & -23:41:03.5 & M \\
		V30       & SR     & 10.941 & 9.074  & 0.962 & 0.211 & 82.50   & 2129.5411 & 18:36:41.06 & -23:58:19.5 & M \\
		V31       & SR     & 11.080 & 9.153  & 0.620 & 0.353 & 95.8253   & 7456.2531 & 18:36:10.12 & -24:05:49.1 & M \\
		V32       & M      & 17.366 & 12.144 & 5.415 & 3.410 & 233.50  & 1774.7109 & 18:35:37.85 & -24:00:02.5 & F \\
		V33       & M      & 15.351 & 10.980 & 4.555 & 2.863 & 251.10  & 7104.4541 & 18:36:14.09 & -24:07:51.0 & F \\
		V34       & SR     & 11.329 & \textit{9.422} & 0.300 & -- & 30.30   & 1731.8218 & 18:36:26.07 & -23:55:33.9 & M \\
		{V35}       & SR     & 11.299 & \textit{9.257}  & 0.387 & 0.153 & 31.90   & 1726.5691 & 18:36:24.04 & -23:54:29.4 & M \\
		V36       & RRab   & 16.757 & 15.804 & 0.798 & 0.531 & 0.626358    & 6479.8038 & 18:35:39.30 & -23:54:06.6 & F \\
		V37       & ECL    & 15.582 & 14.656 & 2.486 & 1.940 & 1.196773    & 7000.1507 & 18:35:48.74 & -23:40:58.9 & F \\
		V38       & RRab   & 15.911 & 15.018 & 1.264 & 0.829 & 0.573788    & 7000.0214 & 18:35:48.95 & -23:42:43.5 & F \\
		V39       & SR     & 16.838 & 12.282 & 0.754 & 0.234 & 66.260   & 1677.8644 & 18:36:09.73 & -23:59:35.6 & F \\
		V40       & SR     & 15.341 & 12.197 & 0.492 & 0.136 & 37.75   & 1674.7791 & 18:36:34.82 & -23:47:23.3 & F \\
		{V41}       & ?      & 17.774 & \textit{12.723} & 0.465 & 0.240 & --    & 1774.7109 & 18:36:58.94 & -23:52:49.1 & F \\
		{V42}       & ?      & 18.286 & \textit{13.128} & 0.614 & 0.334 & --    & 1774.7109 & 18:36:59.59 & -23:47:16.0 & F \\
		{V43}       & ?      & 17.051 & \textit{12.526} & 0.774 & 0.350 & --    & 1774.7109 & 18:36:59.83 & -23:52:19.6 & F \\
		{KT-01}     & EW     & 18.682 & {--}  & 0.273 & -- & 0.313872    & 1785.5321 & 18:36:43.34 & -23:56:25.7 & M \\
		KT-02     & EA/EB  & 17.370 & 16.454 & 0.344 & 0.310 & 0.490630    & 2105.7575 & 18:36:41.82 & -23:56:21.2 & M \\
		KT-03     & EW     & 18.696 & 17.555 & 0.357 & 0.328 & 0.365034    & 1691.9403 & 18:36:41.25 & -23:52:19.9 & F \\
		{KT-04}     & SXPhe  & 16.830 & \textit{16.055}  & 0.210 & -- & 0.035743    & 1692.9396 & 18:36:39.37 & -23:52:25.8 & M \\
		{KT-05}     & SX     & 17.211 & \textit{16.516}  & 0.263 & -- & 0.060833    & 1790.5447 & 18:36:39.08 & -23:50:27.9 & F \\
		KT-07     & EW     & 17.711 & 16.502 & 0.561 & 0.394 & 0.329798    & 1758.4624 & 18:36:36.84 & -23:57:00.7 & F \\
		KT-08     & EW     & 20.193 & 18.844 & 1.012 & 1.280 & 0.363902    & 1765.7125 & 18:36:35.67 & -23:55:15.3 & F \\
		{KT-10}     & SX     & 16.597 & \textit{16.144}  & 0.147 & -- & 0.037499    & 1773.7279 & 18:36:32.30 & -23:54:28.0 & F \\
		KT-12     & RRab   & 16.498 & 15.629 & 1.030 & 0.620 & 0.443610    & 1691.9220 & 18:36:30.93 & -23:53:49.0 & F \\
		KT-13     & EW     & 17.336 & 16.306 & 0.587 & 0.505 & 0.281733    & 1785.6367 & 18:36:30.87 & -23:53:46.3 & M \\
		{KT-14}     & RRc    & 14.175 & \textit{13.329} & 0.333 & 0.222 & 0.374652    & 1786.6719 & 18:36:30.67 & -23:53:54.0 & M \\
		KT-15     & EW     & 16.584 & 15.446 & 0.651 & 0.550 & 0.338112    & 1726.6865 & 18:36:30.11 & -23:49:59.2 & F \\
		KT-16     & RRc    & 14.135 & 13.461 & 0.080 & 0.059 & 0.281901    & 1784.5902 & 18:36:30.36 & -23:57:13.2 & M \\
		KT-18     & EA     & 17.778 & 16.791 & 1.624 & 1.175 & 2.764243    & 1762.5801 & 18:36:29.02 & -23:49:59.3 & F \\
		KT-20     & EW     & 16.618 & 16.003 & 0.229 & 0.203 & 0.288496    & 1786.6430 & 18:36:26.09 & -23:51:26.9 & M \\
		KT-23     & EW     & 16.540 & 15.866 & 0.296 & 0.273 & 0.298524    & 1774.6482 & 18:36:23.84 & -23:51:16.4 & M \\
		{KT-26}     & RRc    & 14.034 & \textit{13.244}  & 0.258 & -- & 0.361362    & 1774.6374 & 18:36:23.16 & -23:53:23.5 & F? \\
		KT-27     & SXPhe  & 16.873 & 16.076 & 0.572 & 0.100 & 0.042174    & 5741.7894 & 18:36:22.54 & -23:55:13.2 & M \\
		{KT-28}     & SXPhe  & 16.153 & \textit{15.309} & 0.098 & 0.174 & 0.055603    & 1792.6850 & 18:36:22.04 & -23:52:06.5 & M \\
		KT-29     & SXPhe     & 16.418 & 15.660 & 0.242 & 0.149 & 0.044325    & 1731.7348 & 18:36:20.96 & -23:55:48.2 & U \\
		{KT-33}     & EW     & 16.983 & --  & 0.111 & -- & 0.244137    & 1773.7339 & 18:36:16.86 & -23:53:54.5 & U \\
		KT-34     & SXPhe  & 16.828 & 15.978 & 0.137 & 0.070 & 0.047318    & 5633.8527 & 18:36:16.89 & -23:55:25.0 & M \\
		KT-36     & RRcBl  & 14.162 & 13.384 & 0.394 & 0.219 & 0.313204    & 1765.6645 & 18:36:15.88 & -23:56:07.1 & M \\
		KT-37     & RRc    & 14.207 & 13.428 & 0.136 & 0.084 & 0.296057    & 6076.7493 & 18:36:13.19 & -23:53:47.0 & M \\
		{KT-38}     & SXPhe  & 16.877 & \textit{16.277}  & 0.092 & -- & 0.032787    & 1692.7132 & 18:36:11.43 & -23:56:47.5 & M \\
		\bottomrule
	\end{tabular}
\end{table*}

\begin{table*}[htbp]
	\addtocounter{table}{-1}
	\Centering
	\caption{Continue}
	\footnotesize
	\begin{tabular}{llrrrrrrlll}
		\toprule
		\textbf{Variable} & \textbf{Type} & \textbf{$<V>$} & \textbf{$<I>$} & \textbf{$A_{V}$} & \textbf{$A_{I}$} & \textbf{Period} & \textbf{$HJD_{max}$} & \textbf{RA} & \textbf{Dec.} & \textbf{Memb.} \\
		&               & (mag)          & (mag)          & (mag)            & (mag)            & (days)          & (d+2450000)          & (J2000.0)   & (J2000.0)     & (M/F/U) \\
		\midrule
		KT-39     & EA     & 17.387 & 16.396 & 0.252 & 0.365 & 1.474834    & 2072.9112 & 18:36:09.99 & -23:51:58.0 & F \\
		KT-40     & EW     & 17.680 & 16.659 & 0.269 & 0.235 & 0.437193    & 2133.6439 & 18:36:08.13 & -23:51:49.2 & F \\
		KT-41     & EW     & 17.958 & 16.606 & 0.723 & 0.710 & 0.293888    & 1732.7333 & 18:36:07.08 & -23:54:13.3 & M \\
		KT-42     & EW     & 17.394 & 16.553 & 0.155 & 0.155 & 0.554883    & 1764.7287 & 18:36:34.65 & -23:52:31.2 & F \\
		KT-43     & EW     & 17.567 & 16.428 & 0.127 & 0.153 & 0.220517    & 1790.7106 & 18:36:24.28 & -23:56:19.0 & M \\
		KT-45     & SXPhe  & 16.591 & 15.806 & 0.298 & 0.183 & 0.050078    & 1762.6514 & 18:36:22.04 & -23:54:41.8 & M \\
		KT-46     & EA     & 19.551 & 18.289 & 1.593 & 1.108 & 0.610198    & 4293.7508 & 18:36:21.75 & -23:58:25.1 & F \\
		{KT-48}     & EW     & 20.403 & --  & 1.478 & -- & 0.338920    & 1773.7182 & 18:36:16.54 & -23:57:36.5 & M \\
		{KT-51}     & sin    & 14.623 & \textit{14.257}  & 0.078 & -- & 2.666559    & 1725.6631 & 18:36:35.08 & -23:53:03.6 & M \\
		KT-54     & SX     & 16.337 & 15.544 & 0.131 & 0.071 & 0.083646    & 1786.5236 & 18:36:24.94 & -23:50:50.2 & F \\
		KT-55     & RRabBl & 14.119 & 13.170 & 1.194 & 0.677 & 0.658739    & 1785.6684 & 18:36:23.24 & -23:53:58.1 & M \\
		{PK-05}     & EW     & 18.401 & --  & 0.372 & -- & 0.242839    & 1758.4793 & 18:36:22.26 & -23:54:32.9 & U \\
		{CV1}       & UG     & 19.398 & --  & 5.794 & -- & --     & 1774.7109 & 18:36:24.66 & -23:54:35.5 & U \\
		{CV2}       & UG     & 19.718 & --  & 2.433 & -- & --    & 1754.4893 & 18:36:02.72 & -23:55:24.6 & F \\
		{P1}        & mlens  & 19.651 & --  & 1.395 & -- & --    & 1758.6032 & 18:36:22.40 & -23:56:29.4 & U \\
		Ku-1      & RRc    & 14.125 & 13.435 & 0.271 & 0.141 & 0.305812    & 1790.6380 & 18:35:59.12 & -23:57:13.4 & M \\
		Ku-2      & RRc    & 14.128 & 13.390 & 0.185 & 0.095 & 0.335236    & 1730.7103 & 18:36:02.97 & -23:50:29.6 & M \\
		Ku-3      & RRc    & 14.045 & 13.320 & 0.463 & 0.224 & 0.334020    & 1692.8589 & 18:36:29.53 & -24:01:33.0 & M \\
		Ku-4      & RRc    & 14.173 & 13.454 & 0.157 & 0.088 & 0.290248    & 4319.5297 & 18:36:31.68 & -23:49:30.5 & M \\
		SLW-4     & SR     & 11.076 & 9.190  & 0.424 & 0.099 & 61.46   & 4327.6176 & 18:36:17.51 & -23:54:26.3 & M \\
		{SLW-5}     & SR     & 11.347 & \textit{9.602}  & 0.167 & -- & 38.00   & 2088.8570 & 18:36:18.38 & -23:54:01.3 & M \\
		{SLW-8}     & SR     & 11.161 & \textit{9.265}  & 0.384 & -- & 51.53   & 2132.6860 & 18:36:21.64 & -23:55:57.0 & M \\
		{SLW-11}    & SR     & 11.367 & \textit{9.574}  & 0.229 & -- & 39.50   & 4326.4739 & 18:36:28.05 & -23:53:23.2 & M \\
		{V102}      & SXPhe  & 16.506 & \textit{15.816}  & 0.046 & -- & 0.024444    & 1772.4849 & 18:36:36.91 & -23:53:47.1 & M \\
		{V103}      & SXPhe  & 16.468 & \textit{15.750}  & 0.146 & -- & 0.034484    & 1730.7039 & 18:36:30.96 & -23:52:52.8 & M \\
		{V104}      & SXPhe     & 16.933 & --  & 0.290 & -- & 0.035295    & 1785.6684 & 18:36:23.86 & -23:54:19.0 & U \\
		{V105}      & SX     & 16.721 & \textit{16.052}  & 0.174 & -- & 0.035436    & 1770.5829 & 18:36:19.88 & -23:55:27.5 & F \\
		{V106}      & SX     & 16.585 & \textit{15.957}  & 0.067 & -- & 0.036600    & 1790.6859 & 18:36:12.43 & -23:51:12.2 & F \\
		{V107}      & SXPhe  & 16.851 & \textit{15.990}  & 0.067 & -- & 0.036672    & 1765.5050 & 18:36:27.17 & -23:55:27.8 & M \\
		{V108}      & SXPhe  & 16.891 & \textit{16.228}  & 0.064 & -- & 0.037069    & 1771.5388 & 18:36:19.03 & -24:00:27.4 & M \\
		{V109}      & SXPhe  & 16.557 & --  & 0.170 & -- & 0.037408    & 1692.7098 & 18:36:03.38 & -23:53:37.4 & M \\
		{V110}      & SXPhe  & 16.504 & \textit{15.805}  & 0.178 & -- & 0.037422    & 1675.8722 & 18:36:13.34 & -23:44:58.6 & M \\
		{V111}      & SXPhe     & 16.591 & --  & 0.430 & -- & 0.049828    & 1773.7241 & 18:36:21.35 & -23:54:38.6 & U \\
		V112      & SXPhe     & 15.943 & 14.858 & 0.460 & 0.204 & 0.062316    & 1773.7310 & 18:36:25.17 & -23:54:01.1 & U \\
		{V113}      & sin    & 15.042 & \textit{13.954}  & 0.034 & -- & 0.134598    & 1772.6591 & 18:36:23.80 & -23:55:05.4 & M \\
		{V114}      & sin    & 19.356 & \textit{18.213}  & 0.279 & -- & 0.138045    & 1772.4824 & 18:36:35.71 & -24:02:05.5 & M \\
		{V115}      & sin    & 19.600 & --  & 0.306 & -- & 0.150652    & 1674.8670 & 18:36:33.23 & -23:59:58.2 & M \\
		{V116}      & susp   & 19.718 & --  & 1.110 & -- & 70.77   & 1770.5959 & 18:36:34.69 & -23:57:52.5 & U \\
		{V117}      & EW?    & 15.850 & \textit{14.704}  & 0.032 & -- & 0.313284    & 1758.7600 & 18:36:16.16 & -23:56:11.1 & M \\
		V118      & EW     & 18.587 & 17.413 & 0.179 & 0.245 & 0.280272    & 2103.7213 & 18:36:06.20 & -23:48:19.6 & M \\
		{V119}      & EW     & 19.502 & --  & 0.337 & -- & 0.308184    & 1671.9334 & 18:36:05.18 & -23:53:14.2 & U \\
		{V120}      & EW     & 17.454 & --  & 0.145 & -- & 0.313017    & 1784.5959 & 18:36:22.29 & -23:54:46.9 & U \\
		{V121}      & EW     & 19.294 & --  & 0.383 & -- & 0.388302    & 1731.6505 & 18:36:14.47 & -23:52:11.0 & U \\
		V122      & EW     & 19.118 & 17.995 & 0.718 & 0.670 & 0.388626    & 1673.8716 & 18:36:30.06 & -23:56:10.6 & U \\
		{V123}      & sin    & 16.996 & \textit{15.954}  & 0.076 & -- & 1.038494    & 1764.5016 & 18:36:34.58 & -23:52:58.4 & M \\
		{V124}      & sin    & 15.873 & --  & 0.039 & -- & 1.040670    & 1765.6694 & 18:36:23.79 & -23:52:24.7 & U \\
		{V125}      & per    & 14.527 & \textit{14.124}  & 0.029 & -- & 0.542888    & 1765.6528 & 18:35:57.48 & -23:47:59.2 & M \\
		{V126}      & sin    & 14.847 & --  & 0.049 & -- & 0.548074    & 4293.7435 & 18:36:23.25 & -23:53:53.6 & M \\
		{V127}      & EW     & 17.342 & \textit{16.606}  & 0.096 & -- & 0.621828    & 1773.6933 & 18:36:55.75 & -23:48:23.9 & M \\
		{V128}      & sin    & 14.553 & \textit{13.234}  & 0.025 & --& 0.881780    & 1649.8941 & 18:36:36.23 & -23:55:21.0 & M \\
		{V129}      & EW     & 15.821 & \textit{15.388}  & 0.066 & -- & 1.394800    & 1692.8836 & 18:36:21.23 & -23:51:37.5 & F \\
		{V130}      & EA/EB  & 16.648 & \textit{15.819}  & 0.078 & -- & 1.445972    & 2063.7813 & 18:36:17.59 & -23:57:14.4 & M \\
		{V131}      & EA     & 15.880 & --  & 0.514 & -- & 1.733622    & 1675.8157 & 18:36:24.18 & -23:54:26.7 & F \\
		{V132}      & susp   & 13.359 & \textit{11.972}  & 0.080 & -- & 1.843108    & 2129.6294 & 18:36:25.73 & -23:56:54.5 & M \\
		{V133}      & EA     & 19.064 & \textit{16.917}  & 0.550 & -- & 2.244228    & 1790.5711 & 18:36:37.30 & -23:53:38.8 & M \\
		V134      & sin    & 17.107 & 16.969 & 0.201 & 0.279 & 2.330917    & 1647.7978 & 18:36:16.63 & -23:55:58.1 & M \\
		{V135}      & EA     & 18.837 & \textit{17.803} & 0.629 & 0.380 & 4.927996    & 7132.3726 & 18:36:55.52 & -23:55:32.3 & M \\
		{V136\_R} & SR     & 15.592 & \textit{14.394}  & 0.095 & -- & 11.7091   & 1677.8500 & 18:36:17.52 & -23:57:31.5 & M \\
		{V136\_f} & V      & 13.926 & \textit{12.618}  & 0.029 & -- & 144.3859  & 4293.7435 & 18:36:39.65 & -23:52:42.7 & M \\
		{V137}      & SR     & 16.041 & --  & 0.245 & -- & 14.97   & 4700.6430 & 18:36:21.69 & -23:53:35.1 & M \\
		{V138}      & SR     & 14.898 & \textit{13.654}  & 0.221 & -- & 34.9127   & 4700.6430 & 18:36:25.51 & -23:52:38.6 & F \\
		{V139}      & SR     & 11.051 & \textit{9.143}  & 0.326 & -- & 64.90   & 4320.5046 & 18:36:15.10 & -23:54:55.0 & M \\
		{V140}      & LPV    & 17.292 & \textit{13.899}  & 1.148 & -- & 3100.00 & 4323.5527 & 18:36:15.38 & -23:52:51.9 & F \\
		{V141}      & SXPhe  & 16.808 & \textit{16.378}  & 0.146 & -- & 0.032720    & 1790.5004 & 18:36:27.00 & -23:49:49.5 & M \\
		{V142}      & EW     & 19.398 & -- & 1.033 & -- & 0.209024    & 1764.6834 & 18:36:30.12 & -23:48:19.7 & M \\
		{V143}      & EW     & 18.233 & \textit{17.531}  & 0.135 & -- & 0.431350    & 1730.8351 & 18:36:24.74 & -23:58:41.3 & M \\

		\bottomrule
	\end{tabular}
\end{table*}

\begin{table*}[htbp]
	\addtocounter{table}{-1}
	\Centering
	\caption{Continue}
	\footnotesize
	\begin{tabular}{llrrrrrrlll}
		\toprule
		\textbf{Variable} & \textbf{Type} & \textbf{$<V>$} & \textbf{$<I>$} & \textbf{$A_{V}$} & \textbf{$A_{I}$} & \textbf{Period} & \textbf{$HJD_{max}$} & \textbf{RA} & \textbf{Dec.} & \textbf{Memb.} \\
		&               & (mag)          & (mag)          & (mag)            & (mag)            & (days)          & (d+2450000)          & (J2000.0)   & (J2000.0)     & (M/F/U) \\
		\midrule
		{V144}      & per    & 17.097 & \textit{15.677}  & 0.201 & -- & 4.572758    & 1758.4624 & 18:36:39.97 & -23:50:54.8 & M \\
		{V145}      & sin    & 17.577 & \textit{16.226}  & 1.614 & -- & 26.2857   & 1694.8946 & 18:36:27.57 & -23:51:21.9 & M \\
		{V146}      & SR     & 11.248 & --  & 0.464 & -- & 48.632   & 1759.7060 & 18:36:10.21 & -23:48:44.7 & M \\
		V147      & SXPhe  & 16.751 & 16.008 & 0.124 & 0.121 & 0.044921    & 7000.0387 & 18:36:31.76 & -23:37:17.3 & M \\
		V148      & SXPhe  & 16.871 & 16.186 & 0.116 & 0.145 & 0.036372    & 7000.0122 & 18:37:11.10 & -23:53:38.5 & M \\
		V149      & ECL    & 16.055 & 15.404 & 0.049 & 0.077 & 0.774442    & 7000.4917 & 18:35:48.81 & -23:50:16.3 & M \\
		{V150}      & EW     & 19.863 & --  & 0.691 & -- & 0.658775    & 1692.7829 & 18:35:57.79 & -23:52:57.1 & M \\
		V151      & ECL    & 16.382 & 15.588 & 0.389 & 0.378 & 0.906254    & 7000.5514 & 18:34:56.03 & -23:53:37.5 & M \\
		V152      & ECL    & 16.681 & 15.580 & 0.497 & 0.471 & 0.350951    & 7000.0344 & 18:37:41.32 & -23:44:06.1 & M \\
		V153      & ECL    & 18.982 & 17.708 & 0.570 & 0.625 & 0.282209    & 7000.1920 & 18:36:26.90 & -23:29:10.5 & M \\
		C29       & EA     & 16.549 & 15.568 & 0.267 & 0.244 & 0.626610    & 1762.5529 & 18:36:15.63 & -23:58:29.2 & M? \\
		\bottomrule
	\end{tabular}
    \center{Columns list the variable identifier, type, intensity-weighted mean $V$ and $I$ magnitudes, amplitudes $A_V$ and $A_I$, period, epoch of maximum light $HJD_\mathrm{max}$, equatorial coordinates (J2000.0), and membership status (M: member, F: field star, U: unclassified). Magnitudes in italics correspond to mean values or data derived from \textit{Gaia}-DR3. Stars labeled with the suffix ``Bl'' in the type column display the Blazhko effect. The complete table is available in the CDS database.}
\end{table*}

\section{Differential reddening correction of M22}
\label{sec:reddening}

As mentioned in Section~\ref{sec:Intro}, M22 is affected by substantial interstellar reddening \citep{Marino2009}. In fact, there is significant evidence for the presence of differential reddening across the field of the cluster. This has the effect of broadening and/or distortioning the evolutionary sequences in the CMD, introducing significant variations in their positions and morphologies. Therefore, if not properly taken into account it may introduce systematic errors in the CMD, potentially leading to wrong estimates of age, metal abundance, and distance. To correct for the differential reddening present in the direction of M22, we applied an iterative procedure fully described in \citet{Pallanca2019,Pallanca2021} and \citet{Cadelano2020}. The same procedure has been adapted and successfully applied to $VI$ photometry of M71 \citep{Cortes2026}. We start from the selection of cluster member stars, at a fixed magnitude range (11.5~$<$~$V$~$<$~20.0), i.e., at the Main Sequence, Sub Giant and Red Giant Branches. We then divided the CMD vertically into magnitude bins of 0.5 mag, except in the region of the Main Sequence-Turn Off and Sub Giant Branches (16.0~$<$~$V$~$<$~17.5), where we used bins of 0.15 mag to provide a finer sampling. For each magnitude bin, we estimated their 3$\sigma$-clipped $V$ and ($V$-$I$) median values to define their representative centroids. Then, we interpolated said centroids in order to create a mean ridge line (MRL), which is used as a reference for estimating the
geometric distance $\Delta X$ of each star in the direction of the reddening vector. The reddening vector is defined using the extinction coefficients $R_{V}$ = 3.1 and  $R_{I}$ = 1.8, computed following \citet{Cardelli1989}. For each likely cluster member, we identified the $n$ nearest reference stars and computed the 3$\sigma$-clipped median of their geometric distances to the MRL, along the reddening vector. This median value corresponds to the star’s assigned displacement $\Delta X$, which
is transformed into the relative differential reddening $\delta E(B-V)$
by using:

\begin{equation}
    \delta E(B-V) = \frac{\Delta X}{\sqrt{R^{2}_{V} + (R_{V}-R_{I})^{2}}}
\end{equation}

To increase the spatial resolution we iteratively performed this
computation three times using the $n$ = 100, 50, and 25 closest stars to each likely cluster member. The differential variations of the color excess within the sampled field of view are significant, ranging between -0.07~$<$~$\delta E(B-V)$~$<$~0.09.

The correction of the magnitudes for differential reddening allowed us to mitigate the broadening of the evolutionary sequences, in particular, it allowed for an improved position of the RR Lyrae stars on the HB, showing a conspicuous segregation between the RRab and the RRc types (see Fig. \ref{Fig:CMD_M22}), typical of Oo II type clusters \citet{Arellano2015, Yepez2020}.

\section{Fourier Analysis of the RR Lyrae Light Curves}
\label{sec:Four}

One of the most useful methods for analyzing RR Lyrae light curves is the Fourier decomposition, a technique described in detail by \citet{Arellano2010}, which provides a thorough explanation of both the methodology and the empirical calibrations. The representation of the light curve in the $V$ filter follows the Fourier series:

\begin{equation}
	m(t) = A_0 + \sum_{k=1}^{N}{A_k \cos\left (\frac{2\pi}{P}~k~(t-E_0) + \phi_k \right) }, 
\end{equation}

\noindent
where $m(t)$ is the magnitude at time $t$, and $P$ y $E_0$ are the period and epoch of maximum, respectively. We derived, for each harmonic, the amplitudes $A_k$ and phases $\phi_k$ as well as the Fourier parameters $\phi_{ij}=j\phi_i-i\phi_j$ and $R_{ij}=A_i/A_j$. 

\subsection{Physical parameters of RR Lyrae stars}
\label{subsec:phy_par}

We can calculate physical parameters such as [Fe/H] and $M_V$ for RRab stars using semi-empirical calibrations from \citet{Jurcsik1996, Nemec2013, Kovacs2001}; and for RRc stars, the calibrations from \citet{Morgan2007, Nemec2013, Kovacs1998}. In the case of [Fe/H] for RRab stars, \citet{Jurcsik1996} calculated this value in their scale ($[\rm{Fe/H}]_J$,  which can be transformed to the Zinn and West (ZW) scale \citep{Zinn1984}, and subsequently converted to the spectroscopic UVES scale using $[\rm{Fe/H}]_{\rm UVES}=-0.413 +0.130[\rm{Fe/H}]_{\rm ZW}-0.356[\rm{Fe/H}]_{\rm ZW}^2$ \citep{Carretta2009}. 

There is a restriction for the use of the calibration of \citet{Jurcsik1996} for the RRab stars; it is only applicable to stars with a deviation parameter $D_m \leq 3.0$. This parameter, defined by these authors (see also \citet{KovacsKanbur1998}), measures the consistency between the morphology of our light curves and the light curves used for [Fe/H] calibration. 

We also calculated $T_{\rm eff}$, $\log (L/L_{\odot})$,  $R/R_{\odot}$ and distance. The calibrations, zero points and technique are described in detail in \citet{Arellano2010}. All physical parameters are listed in Table \ref{tab:par_fis} for member RR Lyrae stars only. The average weighted mean of all physical parameters and their respective standard deviations $\sigma$, are in the last rows. The Fourier coefficients for RR Lyrae members are listed in the Table \ref{tab:fou_coef}.

Three stars were excluded from the mean and $\sigma$ calculations. V18 and KT-36 are Blazhko variables for which no single period adequately fits all the data, making their Fourier decomposition unreliable. KT-37 was excluded due to its anomalously high metallicity ($\rm [Fe/H]_{ZW} = -0.76$), nearly one dex above the cluster mean.

\begin{table*}[htp]
	\footnotesize
	\centering
	\caption{Physical parameters of member RR Lyrae derived from light curve Fourier decomposition.}
	\label{tab:par_fis}
	\begin{tabular}{lccccccccc}
		\hline
		ID & [Fe/H]$_{\rm ZW}$ & [Fe/H]$_{\rm UVES}$ & $M_V$ & $\log T_{\rm eff}$ & $\log(L/L_\odot)$ & $M/M_\odot$ & $R/R_\odot$ & $D$ (kpc) & $D_m$ \\
		\hline
		\multicolumn{10}{c}{RRab}\\
		\hline

		V1    & $-1.73(1)$ & $-1.70(1)$ & $0.49(1)$ & $3.78(1)$ & $1.72(1)$ & $0.91(2)$ & $6.48(1)$ & $3.48(1)$ & 1.6 \\
		V2    & $-1.78(1)$ & $-1.78(1)$ & $0.44(1)$ & $3.81(1)$ & $1.74(1)$ & $0.64(5)$ & $5.80(20)$ & $3.30(1)$ & 2.6 \\
		V4    & $-1.73(1)$ & $-1.71(1)$ & $0.43(1)$ & $3.79(1)$ & $1.74(1)$ & $0.71(2)$ & $6.43(1)$ & $3.35(1)$ & 2.0 \\
		V6    & $-1.72(4)$ & $-1.69(5)$ & $0.42(1)$ & $3.80(1)$ & $1.75(1)$ & $0.98(21)$ & $6.83(60)$ & $3.44(1)$ & 2.9 \\
		V7    & $-1.75(1)$ & $-1.73(1)$ & $0.43(1)$ & $3.80(1)$ & $1.75(1)$ & $0.77(1)$ & $6.27(1)$ & $3.35(1)$ & 1.9 \\
		V10   & $-1.77(1)$ & $-1.76(1)$ & $0.44(1)$ & $3.80(1)$ & $1.74(1)$ & $0.77(1)$ & $6.25(1)$ & $3.35(1)$ & 1.9 \\
		V13   & $-1.79(1)$ & $-1.78(1)$ & $0.42(1)$ & $3.80(1)$ & $1.75(1)$ & $0.74(1)$ & $6.30(1)$ & $3.30(1)$ & 3.1 \\
		V20   & $-1.67(1)$ & $-1.62(1)$ & $0.41(1)$ & $3.79(1)$ & $1.75(1)$ & $0.72(1)$ & $6.68(1)$ & $3.27(1)$ & 1.4 \\
		V23   & $-1.59(1)$ & $-1.52(1)$ & $0.50(1)$ & $3.81(1)$ & $1.71(1)$ & $0.81(10)$ & $5.83(30)$ & $3.48(1)$ & 2.2 \\
		KT-55 & $-1.75(2)$ & $-1.73(2)$ & $0.42(1)$ & $3.80(1)$ & $1.75(1)$ & $0.76(2)$ & $6.33(20)$ & $3.51(1)$ & 2.9 \\
		\hline
		Mean     & $-1.73$ & $-1.70$ & $0.44$ & $3.80$ & $1.74$ & $0.78$ & $6.32$ & $3.38$ & -- \\
		$\sigma$ & $0.06$  & $0.08$  & $0.03$ & $0.01$ & $0.01$ & $0.01$ & $0.32$ & $0.08$ & -- \\
		\hline
		\multicolumn{10}{c}{RRc}\\
		\hline

		V12          & $-1.88(1)$  & $-1.92(1)$  & $0.56(1)$ & $3.86(1)$ & $1.69(1)$ & $0.65(1)$ & $4.62(1)$  & $3.17(1)$ & -- \\
		V15          & $-1.76(3)$  & $-1.74(4)$  & $0.53(1)$ & $3.86(1)$ & $1.70(1)$ & $0.59(1)$ & $4.85(2)$  & $3.14(1)$ & -- \\
		V16          & $-1.82(1)$  & $-1.83(2)$  & $0.55(1)$ & $3.86(1)$ & $1.69(1)$ & $0.60(1)$ & $4.52(3)$  & $3.37(1)$ & -- \\
		V18$^{1}$    & $-1.69(4)$  & $-1.65(5)$  & $0.52(1)$ & $3.86(1)$ & $1.70(1)$ & $0.60(1)$ & $4.51(2)$  & $3.27(1)$ & -- \\
		V19          & $-1.84(1)$  & $-1.85(2)$  & $0.51(1)$ & $3.86(1)$ & $1.71(1)$ & $0.73(1)$ & $5.35(1)$  & $3.26(1)$ & -- \\
		V21          & $-1.81(1)$  & $-1.81(1)$  & $0.55(1)$ & $3.86(1)$ & $1.69(1)$ & $0.63(1)$ & $4.62(1)$  & $3.18(1)$ & -- \\
		V25          & $-1.81(1)$  & $-1.81(1)$  & $0.47(1)$ & $3.86(1)$ & $1.72(1)$ & $0.63(1)$ & $5.20(1)$  & $3.31(1)$ & -- \\
		V27          & $-1.78(10)$ & $-1.78(20)$ & $0.56(1)$ & $3.86(1)$ & $1.69(1)$ & $0.50(1)$ & $4.35(3)$  & $3.13(2)$ & -- \\
		V29          & $-1.56(10)$ & $-1.48(20)$ & $0.61(1)$ & $3.87(1)$ & $1.67(1)$ & $0.60(1)$ & $4.35(2)$  & $3.15(1)$ & -- \\
		KT-14        & $-2.15(1)$  & $-2.33(1)$  & $0.54(1)$ & $3.85(1)$ & $1.70(1)$ & $0.56(3)$ & $4.76(10)$ & $3.30(2)$ & -- \\
		KT-16        & $-1.92(10)$ & $-1.97(10)$ & $0.62(2)$ & $3.86(1)$ & $1.66(1)$ & $0.56(1)$ & $4.04(4)$  & $3.16(3)$ & -- \\
		KT-36$^{1}$  & $-1.53(10)$ & $-1.45(10)$ & $0.57(1)$ & $3.87(1)$ & $1.68(1)$ & $0.68(1)$ & $4.65(2)$  & $3.21(1)$ & -- \\
		KT-37$^{1}$  & $-0.76(30)$ & $-0.72(20)$ & $0.64(1)$ & $3.87(1)$ & $1.65(1)$ & $0.68(1)$ & $4.50(1)$  & $3.17(1)$ & -- \\
		Ku-1         & $-1.51(7)$  & $-1.43(9)$  & $0.62(1)$ & $3.87(1)$ & $1.67(1)$ & $0.57(1)$ & $4.27(1)$  & $3.24(1)$ & -- \\
		Ku-2         & $-1.60(20)$ & $-1.53(20)$ & $0.60(1)$ & $3.86(1)$ & $1.67(1)$ & $0.50(1)$ & $4.26(1)$  & $3.07(1)$ & -- \\
		Ku-3         & $-1.36(3)$  & $-1.24(3)$  & $0.55(1)$ & $3.87(1)$ & $1.69(1)$ & $0.59(2)$ & $4.54(7)$  & $3.18(1)$ & -- \\
		Ku-4         & $-1.38(20)$ & $-1.27(20)$ & $0.67(1)$ & $3.85(1)$ & $1.64(1)$ & $0.50(1)$ & $3.94(2)$  & $2.98(2)$ & -- \\
		\hline
		Mean     & $-1.73$ & $-1.71$ & $0.57$ & $3.86$ & $1.69$ & $0.59$ & $4.55$ & $3.19$ & -- \\
		$\sigma$ & $0.22$  & $0.29$  & $0.05$ & $0.01$ & $0.02$ & $0.06$ & $0.40$ & $0.10$ & -- \\
		\hline
	\end{tabular}
	\center{$^*$ Numbers in parentheses indicate the internal uncertainty expressed to the last digit; e.g. -1.49(3) is equivalent to -1.49$\pm$0.03. $^{1}$ Stars not considered in the mean and $\sigma$ calculations.}
\end{table*}

\begin{table*}[htp]
	\footnotesize
	\centering
	\caption{Fourier coefficients for member RR Lyrae stars in M22.}
	\label{tab:fou_coef}
	\begin{tabular}{lcccccccc}
		\hline
		ID & $A_0$ & $A_1$ & $A_2$ & $A_3$ & $A_4$ & $\varphi_{21}$ & $\varphi_{31}$ & $\varphi_{41}$ \\
		\hline
		\multicolumn{9}{c}{RRab}\\
		\hline
		
		V1    & 14.247(1) & 0.362(1) & 0.181(1) & 0.124(1) & 0.087(1) & 3.912(2) & 1.885(4) & 6.124(5) \\
		V2    & 14.071(1) & 0.370(1) & 0.189(1) & 0.123(1) & 0.086(1) & 4.011(5) & 1.934(7) & 6.259(10) \\
		V4    & 14.113(1) & 0.278(1) & 0.141(1) & 0.088(1) & 0.046(1) & 4.186(4) & 2.289(6) & 0.582(11) \\
		V6    & 14.136(2) & 0.394(3) & 0.202(4) & 0.120(3) & 0.094(4) & 4.017(23) & 1.987(38) & 0.196(48) \\
		V7    & 14.062(1) & 0.377(1) & 0.196(1) & 0.127(1) & 0.089(1) & 4.007(2) & 1.999(3) & 0.114(4) \\
		V10   & 14.115(1) & 0.369(1) & 0.187(1) & 0.122(1) & 0.089(1) & 3.963(4) & 1.964(7) & 0.033(9) \\
		V13   & 14.059(1) & 0.358(1) & 0.190(1) & 0.122(1) & 0.080(1) & 4.078(3) & 2.054(5) & 0.283(7) \\
		V20   & 14.105(1) & 0.242(1) & 0.114(1) & 0.067(1) & 0.029(1) & 4.291(4) & 2.519(6) & 0.864(12) \\
		V23   & 14.279(1) & 0.438(1) & 0.202(1) & 0.142(1) & 0.081(1) & 3.857(7) & 1.782(10) & 5.992(15) \\
		KT-55 & 14.119(1) & 0.372(2) & 0.197(1) & 0.117(2) & 0.070(2) & 4.076(11) & 2.036(18) & 0.163(27) \\
		\hline
		\multicolumn{9}{c}{RRc}\\
		\hline
		
		V12   & 14.179(1) & 0.223(1) & 0.039(1) & 0.016(1) & 0.013(1) & 4.674(7) & 2.654(17) & 1.248(22) \\
		V15   & 14.145(1) & 0.215(1) & 0.023(1) & 0.014(1) & 0.006(1) & 4.980(40) & 3.950(60) & 2.210(150) \\
		V16   & 14.193(1) & 0.213(1) & 0.033(1) & 0.015(1) & 0.013(1) & 4.773(13) & 2.900(28) & 1.571(34) \\
		V18 & 14.131(1) & 0.214(2) & 0.030(2) & 0.017(2) & 0.015(2) & 5.149(51) & 3.194(89) & 2.180(102) \\
		V19   & 14.170(1) & 0.216(1) & 0.018(1) & 0.016(1) & 0.007(1) & 4.963(27) & 3.940(31) & 2.514(68) \\
		V21   & 14.080(1) & 0.211(1) & 0.036(1) & 0.016(1) & 0.014(1) & 4.719(8) & 2.973(18) & 1.696(22) \\
		V25   & 14.172(1) & 0.212(1) & 0.016(1) & 0.015(1) & 0.010(1) & 5.245(21) & 4.282(23) & 2.659(33) \\
		V27   & 13.985(2) & 0.238(3) & 0.030(3) & 0.010(3) & 0.006(3) & 4.984(101) & 3.342(283) & 1.220(491) \\
		V29   & 14.285(1) & 0.200(2) & 0.032(2) & 0.006(2) & 0.004(2) & 4.794(48) & 3.035(249) & 1.612(344) \\
		KT-14 & 14.175(1) & 0.097(1) & 0.004(1) & 0.001(1) & 0.003(1) & 5.064(222) & 2.251(823) & 1.776(317) \\
		KT-16 & 14.135(1) & 0.033(1) & 0.000(1) & 0.000(1) & 0.000(1) & 5.343(530) & 1.366(636) & 1.529(1410) \\
		KT-36 & 14.162(1) & 0.162(2) & 0.019(2) & 0.008(2) & 0.008(2) & 5.003(80) & 3.280(204) & 1.935(197) \\
		KT-37 & 14.207(1) & 0.056(1) & 0.003(1) & 0.000(1) & 0.000(1) & 4.543(46) & 4.081(392) & 0.415(608) \\
		Ku-1  & 14.125(1) & 0.115(1) & 0.011(1) & 0.002(1) & 0.001(1) & 4.842(29) & 3.157(140) & 1.444(230) \\
		Ku-2  & 14.128(1) & 0.073(1) & 0.003(1) & 0.001(1) & 0.000(1) & 4.629(103) & 3.588(340) & 3.241(2707) \\
		Ku-3  & 14.045(1) & 0.194(1) & 0.021(1) & 0.009(1) & 0.009(1) & 4.940(17) & 3.996(40) & 2.478(43) \\
		Ku-4  & 14.173(1) & 0.060(1) & 0.001(1) & 0.001(1) & 0.000(1) & 4.093(263) & 5.027(361) & 0.212(2959) \\
		\hline
	\end{tabular}
	\center{$^*$ Numbers in parentheses is like table \ref{tab:par_fis}.}
\end{table*}

\subsection{Metallicity and distance from \textit{I}-band light curves.}

The calculation of the Physical parameters of RR Lyrae stars via the Fourier decomposition of their $V$-light curves discussed in Section \ref{subsec:phy_par} is a well known and amply discussed approach \citep[e.g.][]{Arellano2024}. The existence of abundant high quality photometry in the Kron-Cousins $I$-band in the OGLE data base \citep{Udalski1992,Udalski2015}, makes it highly desirable to count with specific calibrations to estimate [Fe/H] and the luminosity, and hence the distance, for RR Lyrae stars in globular clusters. Two such calibrations were calculated by \citet{Smolec2005} and with a small modification by \citet{Jurcsik2021}. We shall consider these calibrations and compare the results with those obtained from the $V$-band in section \ref{subsec:phy_par}. This calibration is valid for RRab stars and has the form:

\begin{equation}
	\label{eq:J21}
	{\rm[Fe/H]}J_{\rm ab} = -6.018 -4.261 P+ 1.120 \phi_{31} + 7.466 A_2.
\end{equation}

Independent calibrations for the RRab and RRc stars are those of \citet{Dekany2021}:

\begin{equation}
	\label{eq:D1}
	{\rm[Fe/H]}D_{\rm ab} = -5.819 -6.350 P + 1.248 \phi_{31} + 5.785 A_2,
\end{equation}

\noindent
and

\begin{equation}
	\label{eq:D2}
	{\rm[Fe/H]}D_{\rm c} = -1.821 -10.014 P + 0.325 \phi_{31} -34.704 A_2 +13.835 A_1.
\end{equation}

The Fourier decomposition was performed according to Eq. (1) but for the $I$-band light curves.

To estimate the stellar distance from the I-band light curves, we employed the PMZ calibration	derived by \citet{Prudil_Kunder2024} to calculate $M_I$ (their equation 19);

\begin{equation}
	\label{eq:P24}
	M_I = -1.292~log_{10} P +0.196 {\rm [Fe/H]} +0.197.
\end{equation}

An assumption of the intertellar extinction $R_{VI}$ leads to the distance in parsecs via the equation:

\begin{equation}
	\label{eq:Dist}
	d_I=10^{(1+ 0.2 (I_0-M_I-R_{VI}*E(V-I))}.
\end{equation}

We adopted $R_{VI}=1.205$ from \citet{Prudil2025}.

The resulting [Fe/H]$J_{\rm ab}$ for RRab stars, and [Fe/H]$_D$  $M_V$ and  $D$ (kpc) for cluster member RRab and RRc stars are listed in Table \ref{tab:parfisI}. These results should be compared with the corresponding mean metallicity and distance obtained from the $V$-band light curves decomposition given in Table \ref{tab:par_fis}. As noted by \citet{ArellanoPrudil2026} eq. 3 produces richer iron values richer by about 0.1-0.2 dex and eq. 4 and 5 give poorer values by about 0.3 dex, than results from the $V$-band calibrations after transforming them to the UVES spectroscopic scale of \citet{Carretta2009}. The distance on the other hand, from the PMZ is, within the errors, in very good agreement.

\begin{table}[htp]
	\footnotesize
	\centering
	\caption{Metallicity and distance for member RR Lyrae from $I$-band Fourier light curve decomposition.}
	\begin{tabular}{ccccc}
		\hline
		ID & [Fe/H]$J_{\rm ab}$& [Fe/H]$_D$ & $M_V$ &  $D$ (kpc) \\
		\hline
		\multicolumn{5}{c}{RRab}\\
		\hline
		V1  & -1.473& -2.049 &  0.554 &3.39\\
		V2  & -1.562 &-2.204 &  0.508 &3.40\\
		V4  & -1.450 &-2.116 &  0.500 &3.41\\
		V6  & -1.548 &-2.179 &  0.515 &3.28\\
		V13 & -1.428 &-2.097 &  0.521 &3.29\\
		V20 & -1.426 &-2.111 &  0.488 &3.30\\
		V23 & -1.529 &-1.996 &  0.593 &3.46\\
		\hline
		\multicolumn{5}{c}{RRc}\\
		\hline
		V12 &  -- &-2.008 &  0.653&3.26\\
		V15 &  -- &-2.138 &  0.587&3.30\\
		V16 &  -- &-2.288 &  0.588&3.55\\
		V19 &  -- &-1.732 &  0.671&3.22\\
		V21 &  -- &-2.076 &  0.635&3.23\\
		V25 &  -- &-1.956 &  0.609&3.36\\
		V27 &  -- &-1.707 &  0.705&3.22\\
		V29 &  -- &-1.860 &  0.701&3.37\\
		KT-16 &-- &-3.222 &  0.415&3.70\\
		KT-37 &-- &-2.825 &  0.492&3.53\\
		Ku-1  &-- &-2.563 &  0.542&3.30\\
		Ku-2  &-- &-2.643 &  0.501&3.50\\
		Ku-3  &-- &-2.158 &  0.611&3.35\\
		Ku-4  &-- &-1.221 &  0.856&3.05\\
		\hline
		
		Mean      & $-1.488$ & $-2.150$ & $0.44$ & $3.36$ \\
		$\sigma$ & $0.058$  & $0.416$ & $0.03$ & $0.15$\\
		\hline
		
	\end{tabular}
	
	\label{tab:parfisI}
\end{table}

\subsection{Multiperiodic RRc Stars} 
\label{subsec:multiperiodic_rrc} 

We performed a frequency analysis of the RRc variables in our sample using the software \textsc{Period04} \citep{Lenz2005}. For V15, the period reported in Table \ref{tab:par_fis}, $P = 0.373201$ d, was determined using the string-length method. The frequency analysis with \textsc{Period04} gives a consistent value of $P = 0.3731988$ d. After prewhitening the main pulsation frequency and its harmonics, the residual spectrum reveals an additional independent frequency at $f = 4.35642$ day$^{-1}$, with frequency and amplitude ratios $f_{0.61}/f_\mathrm{1O} = 0.6152$ and $A_{0.61}/A_\mathrm{1O} = 0.061$, respectively. These values are consistent with the so-called RR$_{0.61}$ phenomenon reported in RRc stars in other globular clusters and field stars \citep{Jurcsik2015, Netzel2015a}. To our knowledge, V15 is not included in the OGLE-III Galactic bulge survey analyzed by \citet{Netzel2015a}, since M22 lies outside the OGLE's footprint. Therefore, this constitutes an independent detection of the RR$_{0.61}$ phenomenon in the globular cluster M22. The remaining RRc variables in our sample (V16, V18, KT-36, and Ku-1) were also analyzed with \textsc{Period04}. In all these cases, the phased light curves show clear phase shifts indicative of light curve modulation, probably of Blazhko type. No evidence of double-mode pulsation (RRd) or additional frequencies at the 0.61 frequency ratio was found in any of these stars.

\section{The Color Magnitude Diagram}
\label{sec:cmd}
We searched all point sources in the $Gaia$-DR3 database within the same radius employed by \citet{Vasiliev2021} to guarantee that the majority of stars have a determined probability of membership. The search yields a total of 194,072 point sources in the field of M22, of which 149,308 have magnitudes in $G$, $BP$, and $RP$, necessary for transformation to $V$ and $I$ magnitudes. Finally, only 27,313 have a probability greater than 0.7 of being a member of M22. The final CMD was constructed using only these stars. Subsequently, we cleaned the sample by retaining only the cluster members, resulting in a fairly decontaminated CMD which was then corrected for differential reddening using the methodology mentioned in Sec.\ref{sec:reddening}. 

\begin{figure*}[ht]
	\begin{center}
		\includegraphics[width=16cm]{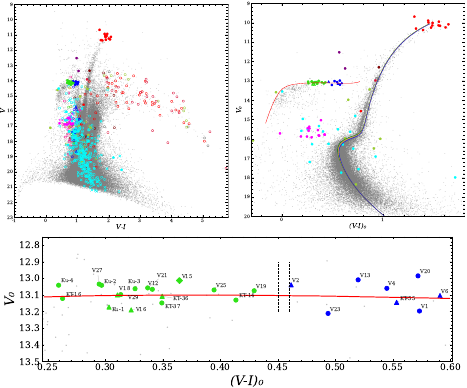}
		\caption{CMD of M22. The left panel displays all stars in the FoV of M22 and all variable stars with $V$ and $I$; the right panel shows the dereddened CMD with member stars and variables only; the bottom panel shows a zoom of the dereddened HB. Displayed on the dereddened CMD are Victoria-Regina isochrones \citep{VandenBerg2014} for ages of 12.0, 12.5, and 13.0 Gyr (green, orange, and blue, respectively, computed for $\mathrm{[Fe/H]} = -1.73$, $Y = 0.27$, and $[\alpha/\mathrm{Fe}] = +0.4$, along with the corresponding ZAHB, all scaled to a distance of 3.27 kpc. In the HB zoom panel, the two vertical dashed lines at $(V-I)_0 = 0.45$ and $0.46$ delimit the first overtone red edge (FORE) of the either-or region as defined by \citet{Arellano2015, Arellano2016}. See Section~\ref{sec:cmd} for the color and symbol scheme.}
		\label{Fig:CMD_M22}
	\end{center}
\end{figure*}

Fig. \ref{Fig:CMD_M22} shows the resulting CMD. RRab Lyrae stars (including Blazhko variables) are shown in blue, while RRc Lyrae stars (including Blazhko and V$_{0.61}$-type) are shown in green. BL Her are plotted in purple. SR are shown in red, whereas Long Period Variables (LPV) and Mira are shown in crimson. SX Phe and related variables are plotted in magenta, while DSCT and SX/DS stars are shown in pink. Eclipsing binaries of all subtypes (EW, EA, EB and their combinations) are shown in cyan. Anomalous Cepheids (AC) are plotted in orange, dwarf novae (UG) in gold, and unclassified periodic variables in yellow-green. Suspected variables are shown in maroon. Member variable stars are indicated with filled symbols, while field variables are shown with open symbols, and variables with uncertain membership are marked with crosses. Blazhko RRab and RRc stars are additionally marked with triangles, and the only V$_{061}$-type RRc star with a diamond.

It is noticed that none of the three isochrones fits is fully satisfactorily. This migh be explained by the chemical complexity of the cluster. \citet{Lee2015} demonstrated that M22 contains two stellar populations, differing not only in their heavy-element abundances but also in their kinematics and spatial distributions, suggesting that M22 most likely formed via a merging event of two globular clusters in a dwarf galaxy environment. \citet{Lee2016} confirmed this bimodal metallicity distribution; who derived a metallicity difference of $\Delta\mathrm{[Fe/H]} \approx 0.20$ dex between the two groups. More recently, \citet{Lee2020} refined this picture by identifying five distinct subpopulations within the two main groups, with the more metal-rich group (G2) showing helium enhancements of $\Delta Y \approx 0.03-0.07$ with respect to its most helium-normal subpopulation. In this context, fitting the CMD of M22 with a single isochrone of fixed metallicity, helium abundance, and age is inherently limited, as no single stellar population model can simultaneously reproduce the full morphology of a metal-complex cluster such as M22.

\section{Bailey diagram and Oosterhoff type}
\label{sec:BD}

The plot of amplitude versus $\log(P)$ for RR Lyrae stars is known as the Bailey diagram. This diagram clearly separates RRc and RRab stars, as they occupy distinct regions with no overlap. The distribution of RRab and RRc stars provides valuable information for determining the Oosterhoff type of a globular cluster, i.e., Oo I or Oo II.

\begin{figure}[ht]
	\begin{center}
		\includegraphics[width=7.5cm]{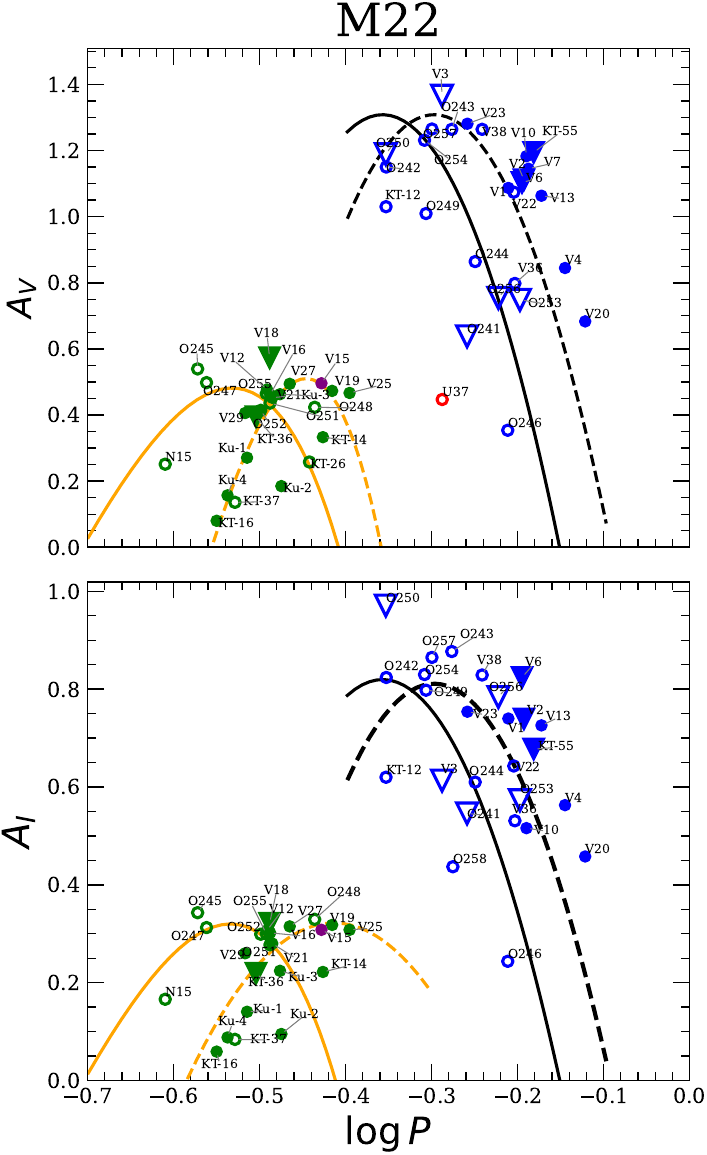}
		\caption{
			Bailey diagram of M22 for $V$ and $I$ amplitudes. Blue and green symbols represent RRab and RRc stars, respectively. Triangles represent stars with Blazhko effect. Purple symbol represents the star V15 discussed in section \ref{subsec:multiperiodic_rrc} and the red symbol represents the star U37 discussed in Appendix \ref{sec: ap-A}. Empty symbols denote field stars. Continuous and segmented loci are the calibrations discussed in section \ref{sec:BD}.
		}
		\label{Fig:Bailey}
	\end{center}
\end{figure}

Figure \ref{Fig:Bailey} shows the Bailey diagram for M22. Based on a study of RRab stars in M3, \cite{Cacciari2005} derived analytical relations for evolved and non-evolved stars, represented by the black dashed and solid lines in the upper panel, respectively. The orange dashed parabola was obtained by \citet{Kunder2013b} for amplitudes in $V$  from a sample of Oo II globular clusters, while both of the orange solid parabolas were derived by \citet{Arellano2015} from the analysis of RRc stars in five Oo I clusters. In the bottom panel, the black solid and dashed lines are from \citet{Kunder2013a} for unevolved and evolved stars, respectively. And the dashed orange line on the lower panel was obtained by \citet{Yepez2020} for Oo II globular clusters.
 
Despite the significant presence of field stars, marked as empty symbols, M22 can be clearly classified as an Oo II cluster, since the confirmed cluster members (filled symbols), align closely with the evolved sequences. It is noticeable that Blazhko variables do follow the evolved Oo II sequences. In the HB zoom of Fig. \ref{Fig:CMD_M22} we can appreciate the segregation between the RRab and RRc stars; both groups are separated by the first overtone red edge (FORE) calculated by \citet{Arellano2015, Arellano2016}.

The clean pulsation mode splitting is an inevitable characteristic in all Oo II clusters according to  \citet{Arellano2015,Arellano2024b} and \citet{Yepez2020}. Also consistent with the classification as an Oo II is the Period average of all members RRab of 0.65$\pm$0.05 days.

\section{Distance of M22 from SX Phe P-L relations}
\label{sec:SXPL}

\begin{figure}[htbp]
	\begin{center}
		\includegraphics[width=8.5cm]{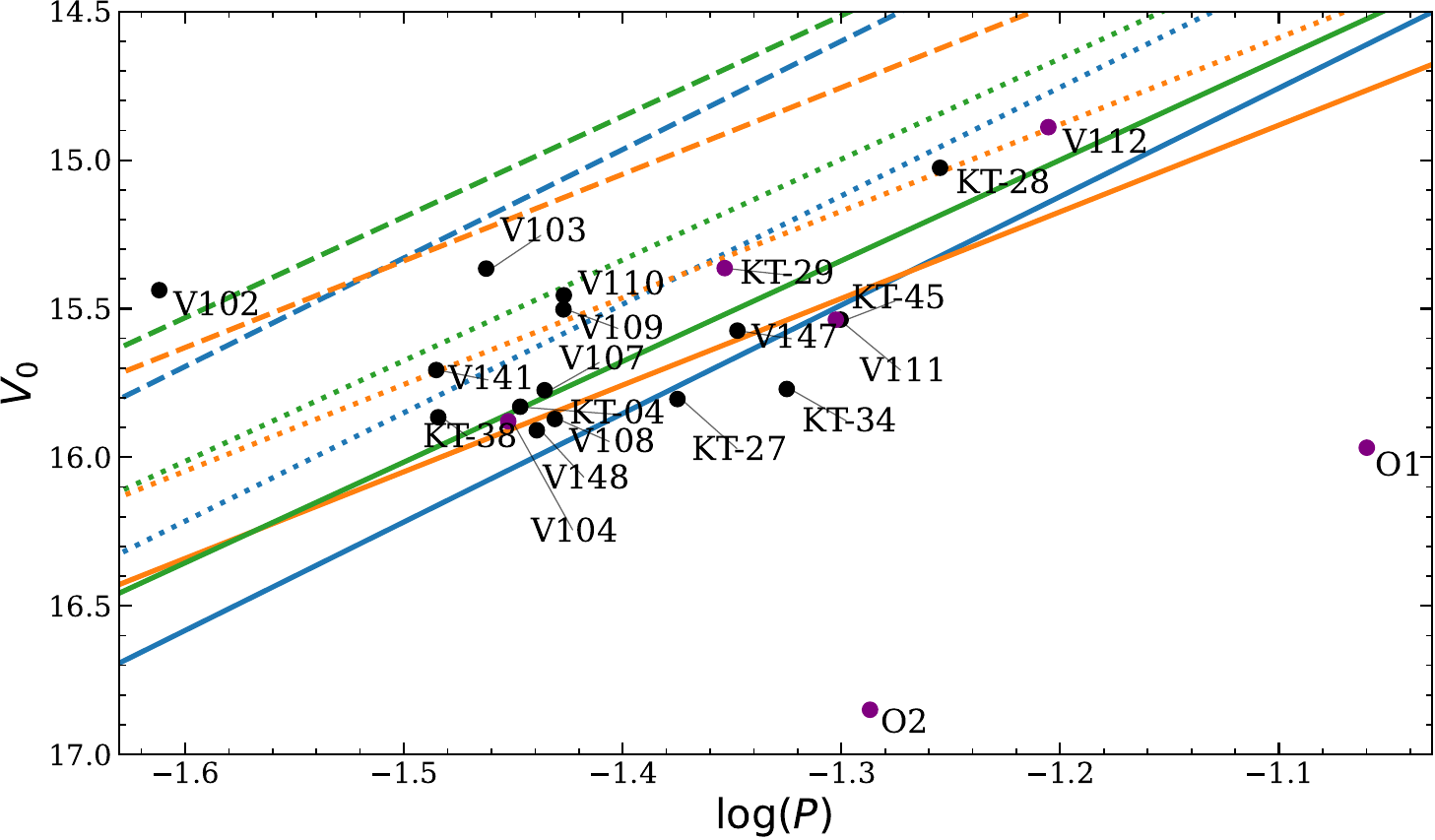}
		\caption{P-L relations of SX Phe stars. Blue, orange, and green lines correspond to the relations of \citet{Poretti2008}, \citet{Arellano2011} and \citet{CohenSara2012}, respectively. Solid, dotted, and dashed lines indicate the fundamental, first overtone, and second overtone modes. Black and purple circles are cluster members and unclassified stars, respectively.}
		\label{Fig:SX_PL}
	\end{center}
\end{figure}

An independent estimation of the distance to M22 can be obtained from the period-luminosity (P-L) relation for SX Phe stars. In the field of M22 there are 45 stars identified as SX Phe or $\delta$ Scuti, of which 24 are classified as field stars, 15 as members, and the remainder as unclassified; the periods of all of them range between 0.024 and 0.062 d. Excluding the field stars, we show in Fig. \ref{Fig:SX_PL} the $V_0$ versus $\log(P)$ diagram for the remaining 21 stars. The dereddened magnitudes $V_0$ were computed using individual reddening values for each star, obtained from our previously discussed differential reddening analysis. The P-L relations of \citet{Poretti2008}, \citet{Arellano2011}, and \citet{CohenSara2012} were scaled to the distance of 3.27 kpc derived from the RR Lyrae stars, and are shown for the fundamental mode (FM), first overtone (1O), and second overtone (2O) as solid, dotted, and dashed lines, respectively.

For each star, the pulsation mode was assigned according to its position in the $V_0$ vs. $\log(P)$ diagram, whether FM, 1O, or 2O. Based on this assignment, we calculated the individual distance of each of the 15 confirmed member stars, obtaining mean distances of $3.30 \pm 0.18$, $3.47 \pm 0.21$, and $3.22 \pm 0.21$ kpc from the calibrations of \citet{Arellano2011}, \citet{CohenSara2012}, and \citet{Poretti2008}, respectively, in good agreement with the distance obtained from the RR Lyrae stars. The resulting individual distances are listed in Table \ref{tab:SX_dist}. Among the variables labeled with membership U, also plotted in Fig. \ref{Fig:SX_PL} for comparison, four stars (KT-29, V104, V111, and V112) fall within the locus defined by the P-L relations; if they were members of the cluster, their individual distances, also listed in Table \ref{tab:SX_dist}, would be consistent with the adopted distance. While the P-L position alone is insufficient evidence to confirm their membership, these stars warrant further investigation. In contrast, O1 and O2 fall far from these relations, leading us to think that they are not members of M22.

\begin{table}[htbp]
	\footnotesize
	\centering
	\caption{Distance estimates for SX Phoenicis stars in M22.}
    \label{tab:SX_dist}
	\begin{tabular}{lccccc}
		\hline
		ID & $E(B-V)$ & Mode & $D_{\rm AF2011}$ & $D_{\rm CS2012}$ & $D_{\rm Por2008}$ \\
		& & & (kpc) & (kpc) & (kpc) \\
		\hline
		KT-04 & 0.3227 & FM & 3.17 & 3.26 & 2.99 \\
		KT-27 & 0.3448 & FM & 3.46 & 3.60 & 3.33 \\
		KT-28 & 0.3636 & 1O & 3.27 & 3.59 & 3.41 \\
		KT-34 & 0.3412 & FM & 3.64 & 3.84 & 3.57 \\
		KT-38 & 0.3263 & 1O & 3.54 & 3.69 & 3.41 \\
		KT-45 & 0.34 & FM & 3.38 & 3.58 & 3.34 \\
		V102  & 0.3447 & 2O & 2.94 & 3.08 & 2.85 \\
		V103  & 0.3559 & 2O & 3.48 & 3.76 & 3.54 \\
		V107  & 0.3472 & FM & 3.14 & 3.23 & 2.97 \\
		V108  & 0.3288 & FM & 3.31 & 3.41 & 3.13 \\
		V109  & 0.34 & 1O & 3.23 & 3.41 & 3.18 \\
		V110  & 0.3383 & 1O & 3.17 & 3.34 & 3.11 \\
		V141  & 0.3554 & 1O & 3.29 & 3.42 & 3.17 \\
		V147  & 0.3796 & FM & 3.23 & 3.38 & 3.14 \\
		V148  & 0.3102 & FM & 3.33 & 3.42 & 3.14 \\
		KT-29$^*$ & 0.34 & 1O & 3.35 & 3.59 & 3.38 \\
		V104$^*$  & 0.34 & FM & 3.22 & 3.31 & 3.03 \\
		V111$^*$  & 0.34 & FM & 3.37 & 3.57 & 3.33 \\
		V112$^*$  & 0.34 & 1O & 3.28 & 3.64 & 3.48 \\
		\hline
		Mean     & -- & -- & 3.30 & 3.47 & 3.22 \\
		$\sigma$ & -- & -- & 0.18 & 0.21 & 0.21 \\
		\hline
	\end{tabular}
    \center{$^*$ Not considered in the mean and $\sigma$ calculations.}
\end{table}

\section{Conclusions}
\label{sec:conclusions}
We identified 94 variable stars as members of M22, including 10 RRab, 17 RRc, 2 Type II Cepheids, 15 SX Phe, 15 SR stars, 22 eclipsing binaries and 13 of other types, from a total of 666 variables in the cluster's field of view. This membership classification was established by cross-matching our catalogue against five independent analyses (R17, R18, \citealt{Vasiliev2021}, \citealt{Alonso-Garcia2021}, and \citealt{Prudil2024}), constituting a homogenized compilation of variable star membership in M22. Among these 94 members, 7 were recovered from stars previously labeled as unclassified by R17, 3 are new members identified from the OGLE database, and 4 are new members from \citet{Alonso-Garcia2025}; none of these 14 stars are included in the Catalogue of Variable Stars in Globular Clusters (CVSGC, \citealt{Clement2001}).

We performed a Fourier decomposition of the $V$ light curves of the member RR Lyrae stars to estimate the mean metallicity and distance to M22. From the combined sample of 24 RRab and RRc stars, we obtained $\mathrm{[Fe/H]} = -1.73 \pm 0.17$ dex and a distance $d = 3.27 \pm 0.14$ kpc. Our derived [Fe/H] is consistent with recent high-precision estimates by \citet{McKenzie2022}, which report a [Fe/H] range of -1.973 to -1.696 dex between populations. Regarding distance estimates, our result is in good agreement with independent estimates of $d \approx 3.3$ kpc \citep{Baumgardt2021}, the photometric estimate of $d = 3.2 \pm 0.2$ kpc obtained by \cite{Kunder2013b} from the cluster RR Lyrae population, and the $Gaia$-DR2-based distance of $d = 3.24 \pm 0.08$ kpc derived by \cite{Baumgardt2019}. A similar value was also obtained by \cite{Alonso-Garcia2021}, who reported $d = 3.20 \pm 0.06$ kpc through a recalibration of the period–luminosity relations anchored to the distance estimate of \cite{Baumgardt2019}.

We also derived the distance from the $I$-band light curves, obtaining $d = 3.36 \pm 0.15$ kpc, in good agreement with the value obtained in $V$; however, the metallicity derived in the $I$ band does show systematic differences with respect to that in $V$, depending on the calibration used.

The distances derived from the SX Phe stars range from $3.22 \pm 0.21$ to $3.47 \pm 0.21$ kpc, depending on the calibration used, and remain consistent with the distance obtained from the RR Lyrae stars.

The Bailey diagram, the member RRab period average of $0.65 \pm 0.05$ d, and the mode splitting in the HB confirm M22 as an Oosterhoff Type II (OoII) globular cluster, as the distribution of the RR Lyrae stars matches the corresponding relationship for this type. Remarkably, the RR Lyrae stars exhibiting the Blazhko effect also follow these relations.

We detected the $f_{0.61}$ phenomenon in the RRc star V15, with frequency and amplitude ratios consistent with values previously reported in other globular clusters and field stars \citep{Jurcsik2015, Netzel2015a}. The member stars V16, V18, and KT-36 exhibit light curves with the Blazhko effect.

The CMD was differentially dereddened, and almost all member variables lie in their expected regions according to their types; however, none of the plotted isochrones provided a satisfactory fit. This can be explained by the chemical complexity of M22, which hosts at least two stellar populations with distinct metallicities.

\renewcommand{\refname}{REFERENCES}
\bibliography{M22.bib}

%----------------------------------------------------------

\section{ACKNOWLEDGEMENTS}

AAF thanks DGAPA-UNAM for financial support through the IN103024.project. 

This work has made use of data from the European Space Agency (ESA) mission
{\it Gaia} (\url{https://www.cosmos.esa.int/gaia}), processed by the {\it Gaia}
Data Processing and Analysis Consortium (DPAC,
\url{https://www.cosmos.esa.int/web/gaia/dpac/consortium}). Funding for the DPAC
has been provided by national institutions, in particular the institutions
participating in the {\it Gaia} Multilateral Agreement.

\begin{appendices}

\section{Comments about individual stars}
\label{sec: ap-A}

\subsection{RR Lyrae stars}
\begin{description}

\item[\textbf{V3, V10, KT-14, KT-55.}] These stars only have \textit{Gaia}-DR3 data, and an adequate fit could not be performed due to the lack of points; therefore, their position in the CMD may be incorrect.

\item[\textbf{V7.}] This star lacks data in $I$, so its light curve is incomplete, preventing the determination of $\langle I \rangle$, and therefore its color needed for its proper location in the CMD.

\item [\textbf{V21.}] This star is reclassified as an RRc variable, in agreement with \citet{Alonso-Garcia2025}.

\item[\textbf{V27, V29.}] Despite being classified as members by \citet{Prudil2024}, a note in the CVSGC indicates that their membership may still be uncertain. However, the derived metallicity and distance are consistent with cluster membership.

\item[\textbf{KT-26.}] This star lacks data in $I$ and is classified as a field star according to \citet{Vasiliev2021}, although R17 consider it a cluster member. Its metallicity $[\mathrm{Fe/H}] = -2.0$ is lower than the cluster average, but its distance of 2.93 kpc suggests that it is currently immersed in the cluster.

\item[\textbf{KT-36.}] There is an offset between its filters, and no period could be found that simultaneously fitted all the data. The star likely shows evidence of a period change over time; an $O-C$ diagram would be needed to confirm this.

\item[\textbf{KT-37.}] This is the member RR Lyrae star with the highest metallicity in the sample. It is likely not a true member and may instead be a field star with a proper motion similar to the cluster.

\item[\textbf{N15.}] There is a discrepancy in the classification of this star: the OGLE database identifies it as an RR Lyrae star, whereas \citet{Rozyczka2017} classify it as a Delta Scuti star. Based on our analysis, we conclude that this star is an RR Lyrae variable.

\item[\textbf{O239.}] The maximum may be poorly determined due to the lack of data; the light curve in $I$ suggests a probable Blazhko effect.

\item[\textbf{O240, O241.}] These are RRab stars without proper motion data. However, the distance derived from their light curve places them at more than 10 kpc, indicating that they are field stars. 

\item[\textbf{O248.}] This star is now classified as a Blazhko variable. Its amplitude in $I$ is unusually large compared to what is expected for its period, while in $V$ the amplitude falls within the expected range in the Bailey diagram.

\item[\textbf{O251, O254.}] These stars are now classified as Blazhko variables.

\end{description}

\subsection{SX Phe stars}

\begin{description}
\item[\textbf{KT-29, V104, V111, V112.}] Although classified as non-members by R17, they could be members according to the period-luminosity relation of SX Phe stars.

\item[\textbf{KT-27.}] A period of 0.042174405 d is required for a proper phased light curve, with all decimal places. The same applies to the following stars: KT-28: 0.055602946 d; KT-29: 0.044324985 d; KT-34: 0.0473177775 d; KT-45: 0.05007783 d; KT-54: 0.08364642 d; V112: 0.062316127 d.

\item[\textbf{O2.}] This star has two active periods: 0.051652 d and 0.066029 d.

\end{description}

\subsection{Type II Cepheids}
\begin{description}
\item[\textbf{V11 and V24.}] Both stars are classified as CW in the CVSGC, but V24 is classified as BL Her star in R17. Based on their periods, we classify both stars as BL Her stars.
\end{description}

\subsection{SR and LPV stars}

\begin{description}
\item[\textbf{V8.}] This star lacks data in $I$, so its light curve is incomplete, preventing the determination of $\langle I \rangle$ and therefore its color needed for its proper location in the CMD.

\item[\textbf{V136\_R.}] In CASE data, V136 appears with the same designation in R17 and R18 but with different coordinates; we refer to these as V136\_R and V136\_f, respectively. This star is a member according to the membership analysis, but it is too faint to be an SR. If it were a true SR, it could be a field star with a proper motion similar to that of the cluster.

\item[\textbf{V136\_f.}] This star is a member according to \citet{Vasiliev2021}. Based on its period, it could be classified as an SR, but its magnitude in $V$ places it below the tip of the RGB. Data in $I$ are needed to confirm its position in the CMD and provide a more reliable classification; it has been decided to keep it as an LPV.

\item[\textbf{N176, N177, N178, N184, N185, N187, N188, N189, N191, N192,}]
\item[\textbf{N195, N198, N199, N201, N203.}] These stars were previously classified as suspected variables by R2017. They exhibit variability with periods ranging from tens to hundreds of days, suggesting they could be SR stars; however, none are cluster members. Due to the lack of a more detailed analysis and the difficulty in obtaining a satisfactory period fit for most of them, they have been provisionally classified as LPV stars.

\end{description}

\subsection{Eclipsing binaries and other variables}

\begin{description}

\item[\textbf{V41-V43.}] These stars only have \textit{Gaia}-DR3 data; when plotted against HJD no further analysis can be performed. More data is needed.

\item[\textbf{KT-51.}] This is a very low amplitude variable; the variability is not clearly visible.

\item[\textbf{V125.}] This star is now classified as an eclipsing binary.

\item[\textbf{U37.}] R17 classified this star as a likely RR Lyrae star. With the purpose of clarifying its nature, we added it to the Bailey Diagram (Fig. \ref{Fig:Bailey}, top panel) to check if its position aligns with the loci occupied by either the RRab or RRc stars. Nevertheless, U37 falls right between both distributions. With a period of 0.515592 d the star is likely an RRab star. However, its amplitude seems too small even for an OoI type RR Lyrae (Fig. \ref{Fig:U37}). The star is nearly 5 magnitudes fainter than the cluster HB and therefore, being a RR Lyrae, it is a field star behind the cluster. The available light curve is very scattered since measuring this faint star is uncertain, and it is not unlikely that the star is blended with a brighter star in the cluster field. These circumstances might contribute to its apparent small amplitude.

\begin{figure}[ht]
	\begin{center}
		\includegraphics[width=7.5cm]{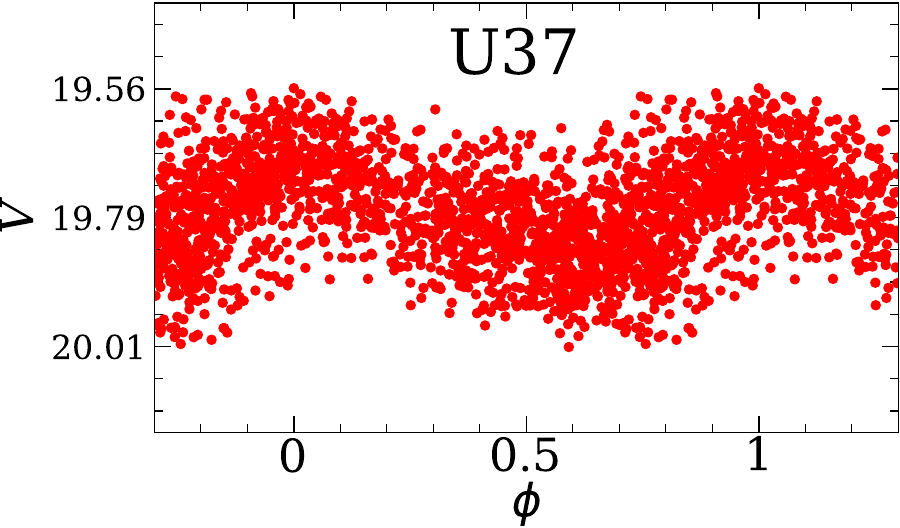}
		\caption{V light curve of U37.
		}
		\label{Fig:U37}
	\end{center}
\end{figure} 

\item[\textbf{U55.}] This is an eclipsing binary. Using a period of 4.492702 d, the light curve shows two eclipses, one deeper than the other.

\item[\textbf{N16.}] R17 yields a period of 0.258814 d, which suggests an eclipsing binary; \textsc{Period04} gives a period of 0.1294079 d, which is practically half. However, since no frequency is associated with the R17 period, the \textsc{Period04} value was adopted.

\item[\textbf{N107.}] This is a Delta Scuti star; the period found corresponds to the main frequency detected with \textsc{Period04}.

\item[\textbf{N126.}] This is an eclipsing binary; both eclipses are clearly visible with a period of 7.906869 d. 

\item[\textbf{N155.}] The maximum does not coincide between the different data groups, and the amplitude in the \textit{Gaia}-DR3 data is too large. The period is likely incorrect; no better period could be found.

\item[\textbf{N182.}] This star displays a low-amplitude light curve; however, a series of points deviates from the general trend. A more detailed analysis is required to better characterize its variability.

\item[\textbf{V150.}] The magnitude reported by R17 (19.73) and that from \textit{Gaia}-DR3 (18.83) are inconsistent, differing 
by nearly one magnitude. Data in $I$ are also lacking to confirm its position in the CMD and provide further information.

\end{description}

\subsection{Membership notes}
\begin{description}
\item[\textbf{KT-01, KT-13, KT-41, KT-48, V126, V137.}] These stars appear as members in \citet{Prudil2024}, but no match could be found with the \citet{Vasiliev2021} data. The membership from \citet{Prudil2024} was adopted.

\item[\textbf{KT-08, CV2 and V131.}] These stars are classified as non-members according to \citet{Prudil2024}, but have no data in \citet{Vasiliev2021}.
\end{description}

\end{appendices}

\end{document}